\documentclass[iop]{emulateapj}

\slugcomment{}

\shorttitle{Star formation around mid-infrared bubble N37}
\shortauthors{T. Baug et al.}
\begin{document}

\title{Star formation around mid-infrared bubble N37: Evidence of cloud-cloud collision}
\author{T. Baug}
\affil{Department of Astronomy and Astrophysics, Tata Institute of Fundamental Research, Homi Bhabha Road, Mumbai-400005, India}

\author{L.K. Dewangan}
\affil{Physical Research Laboratory, Navrangpura, Ahmedabad - 380 009, India.}

\author{D.K. Ojha}
\affil{Department of Astronomy and Astrophysics, Tata Institute of Fundamental Research, Homi Bhabha Road, Mumbai-400005, India}

\and
\author{J.P. Ninan}
\affil{Department of Astronomy and Astrophysics, Tata Institute of Fundamental Research, Homi Bhabha Road, Mumbai-400005, India}

\email{tapas.baug@tifr.res.in}
%\altaffiltext{1}{Department of Astronomy and Astrophysics, Tata Institute of Fundamental Research, Homi Bhabha Road, Mumbai 400 005, India.}
%\altaffiltext{2}{Physical Research Laboratory, Navrangpura, Ahmedabad - 380 009, India.}

\begin{abstract}
We have performed a multi-wavelength analysis of a mid-infrared (MIR) bubble N37 and its surrounding environment. The selected 15$' \times$15$'$ area around the
 bubble contains two molecular clouds (N37 cloud; V$_{lsr}\sim$37--43 km s$^{-1}$,  and C25.29+0.31; V$_{lsr}\sim$43--48 km s$^{-1}$) along the line of
 sight. A total of seven OB stars are identified towards the bubble N37 using photometric criteria, and two of them are spectroscopically confirmed as O9V
 and B0V stars. Spectro-photometric distances of these two sources confirm their physical association with the bubble. The O9V star is appeared to be the primary
 ionizing source of the region, which is also in agreement with the desired Lyman continuum flux analysis estimated from the 20 cm data. The presence of the expanding H{\sc ii}
 region is revealed in the N37 cloud which could be responsible for the MIR bubble. Using the $^{13}$CO line data and photometric data, several cold molecular
 condensations as well as clusters of young stellar objects (YSOs) are identified in the N37 cloud, revealing ongoing star formation (SF) activities. However, the
 analysis of ages of YSOs and the dynamical age of the H{\sc ii} region do not support the origin of SF due to the influence of OB stars. The position-velocity
 analysis of $^{13}$CO data reveals that two molecular clouds are inter-connected by a bridge-like structure, favoring the onset of a cloud-cloud collision
 process. The SF activities (i.e. the formation of YSOs clusters and OB stars) in the N37 cloud are possibly influenced by the cloud-cloud collision.
\end{abstract}
\section{Introduction}
Massive stars ($>$8 M$_\odot$) play a crucial role in the evolution of their host galaxies, but their exact formation and evolution mechanisms are 
 still under debate \citep{zinnecker07,peters12,dale15,kuiper15}. It is not yet understood whether the formation of massive stars is only
 a scaled-up version of birth process of low mass stars, or is it a completely different process. One can find more details about the current theoretical
 scenarios of massive star formation in the recent reviews by \citet{zinnecker07} and \citet{tan14}. Recently, a collision between two molecular clouds
 followed by a strong shock compression of gas is considered as a probable formation mechanism of massive stars \citep{furukawa09,ohama10,fukui14,torii15}. 
 \citet{habe92} numerically found that the head-on collision between two non-identical molecular clouds can trigger the formation of massive stars, and
 such process could also form a broken bubble-like structure. In a detailed study of RCW 120 star-forming region using the molecular line data,
 \citet{torii15} reported that the collision between two nearby molecular clouds  has triggered the formation of an O star in RCW 120 in a short time
 scale. However, observational evidences for the formation of O stars via a collision between two molecular clouds are still very rare.
 
Massive stars can significantly influence the surrounding interstellar medium (ISM) through their energetics such as ionizing radiation, stellar winds,
 and radiation pressure. They have an ability to help in accumulation of surrounding materials (i.e., positive feedback) and/or to disperse matter into the
 ISM. Furthermore, they can also affect the star formation positively and negatively \citep{deharveng10}. The positive feedback of massive stars can trigger
 the birth of a new generation of stars including young massive star(s). More details about the various processes of triggered star formation can be found
 in the review article by \citet{elmegreen98}. However, the feedback processes of massive stars are not yet well understood, and the direct observational
 proof of triggered star formation by massive stars is rare. But the influence of massive stars on their surroundings can be studied with several other
 observational signatures (like H {\sc ii} region, wind-blown or radiation driven Galactic bubble, etc.).

Recently, {\it Spitzer} observations have revealed thousands of ring/shell/bubble-like structures in the 8 $\mu$m images \citep{churchwell06,churchwell07,simpson12},
 and many of them often enclose the H\,{\sc ii} regions. Hence, the bubbles associated with H\,{\sc ii} regions are potential targets to probe the physical
 processes governing the interaction and feedback effect of massive stars on their surroundings. Additionally, these sites are often grouped with the
 infrared dark clouds (IRDCs) and young stellar clusters, which also allow to understand the formation and evolution of these stellar clusters. 
 
 In this paper, we present a multi-wavelength study of such a mid-infrared (MIR) bubble, N37 \citep[$l=$ 25$^\circ$.292, $b=$ 0$^\circ$.293;][]{churchwell06},
 which is associated with an H\,{\sc ii} region, G025.292+00.293 \citep{churchwell06,deharveng10,beaumont10}. The bubble N37 is classified as a broken
 or incomplete ring with an average radius and thickness of 1$\farcm$77 and 0$\farcm$49, respectively \citep{churchwell06}. The bubble is found in the direction
 of the H\,{\sc ii} region RCW 173 (Sh2-60) \citep[see Figure~9 in][]{marco11}. The velocity of the ionized gas \citep[$\sim$39.6~km\,s$^{-1}$;][]{hou14} is in
 agreement with the line-of-sight velocity of the molecular gas \citep[$\sim$41~km\,s$^{-1}$;][]{beaumont10,shirley13} towards the bubble N37, indicating the
 physical association of the ionized and molecular emissions. Presence of several IRDCs are also reported around the N37 bubble by \citet{peretto09}. \citet{marco11}
 analyzed the photometry and spectroscopy of stars in the direction of the H\,{\sc ii} region, RCW 173, and found that most of the stars in
 the field are reddened B-type stars. They also identified a star a805 (G025.2465+00.3011) having spectral type of O7II and suggested this as
 the main ionizing source in the area. Several kinematic distances (2.6, 3.1, 3.3, 12.3, and 12.6 kpc) to the region are listed in
 the literature \citep[e.g.][]{beaumont10,churchwell06,blitz82,watson10,deharveng10}. However, it has been pointed out by \citet{churchwell06} that the
 MIR bubbles located at the Galactic plane are likely to be veiled behind the foreground diffused emission if they are situated at a distance larger than
 $\sim$8 kpc. Hence, it is unlikely for the bubble N37 to be located at a distance of about 12 kpc. Therefore, in this work, we have adopted a distance 
 of 3.0 kpc, the average value of all available near-kinematic distance estimates.

We infer from the previous studies that the bubble is associated with an H\,{\sc ii} region and an IRDC together. However, the physical conditions inside
 and around the bubble N37 are not yet known, and the ionizing source(s) of the bubble is yet to be identified. Furthermore, the impact of the energetics of
 massive star(s) on its local environment is not yet explored. The detailed multi-wavelength study of the region will allow us to study the ongoing physical
 processes within and around the bubble N37. To study the physical environment and star formation mechanisms around the bubble, we employ multi-wavelength data
 covering from the optical, near-infrared (NIR) to radio wavelengths.

The paper is presented in the following way. In Section~\ref{sec:observations}, we describe the details of the multi-wavelength data.
 We discuss the overall morphology of the region in Section~\ref{sec:morphology}. In Section~\ref{sec:result}, we present the main results of our analysis. The
 possible star formation scenarios based on the multi-wavelength outcomes are discussed in Section~\ref{sec:discussion}. Finally, we conclude in Section~\ref{sec:conclusions}.
\section{Observations and data reduction}
\label{sec:observations}
\begin{figure*} 
\epsscale{0.55}
\plotone{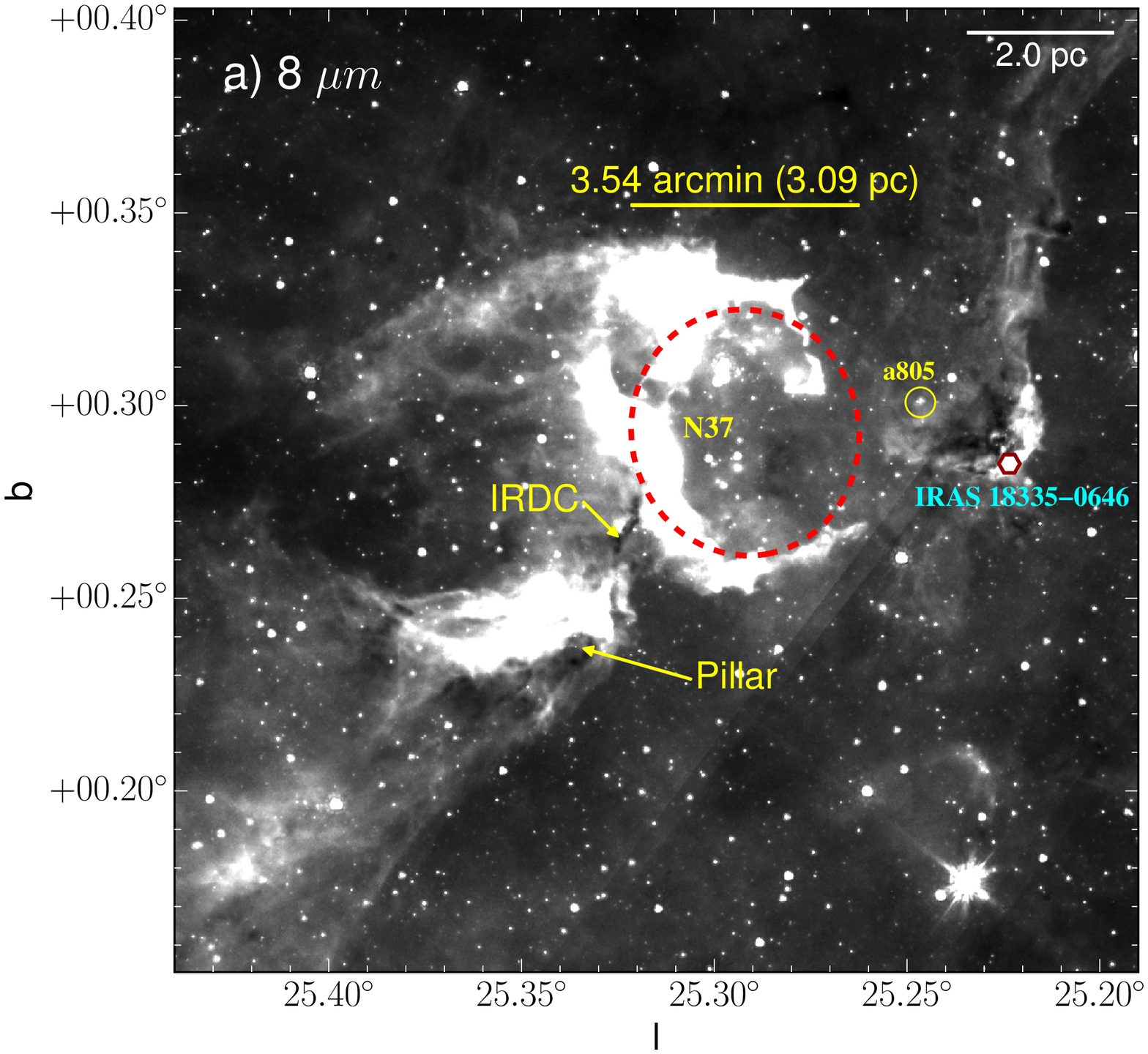}
\epsscale{0.55}
\plotone{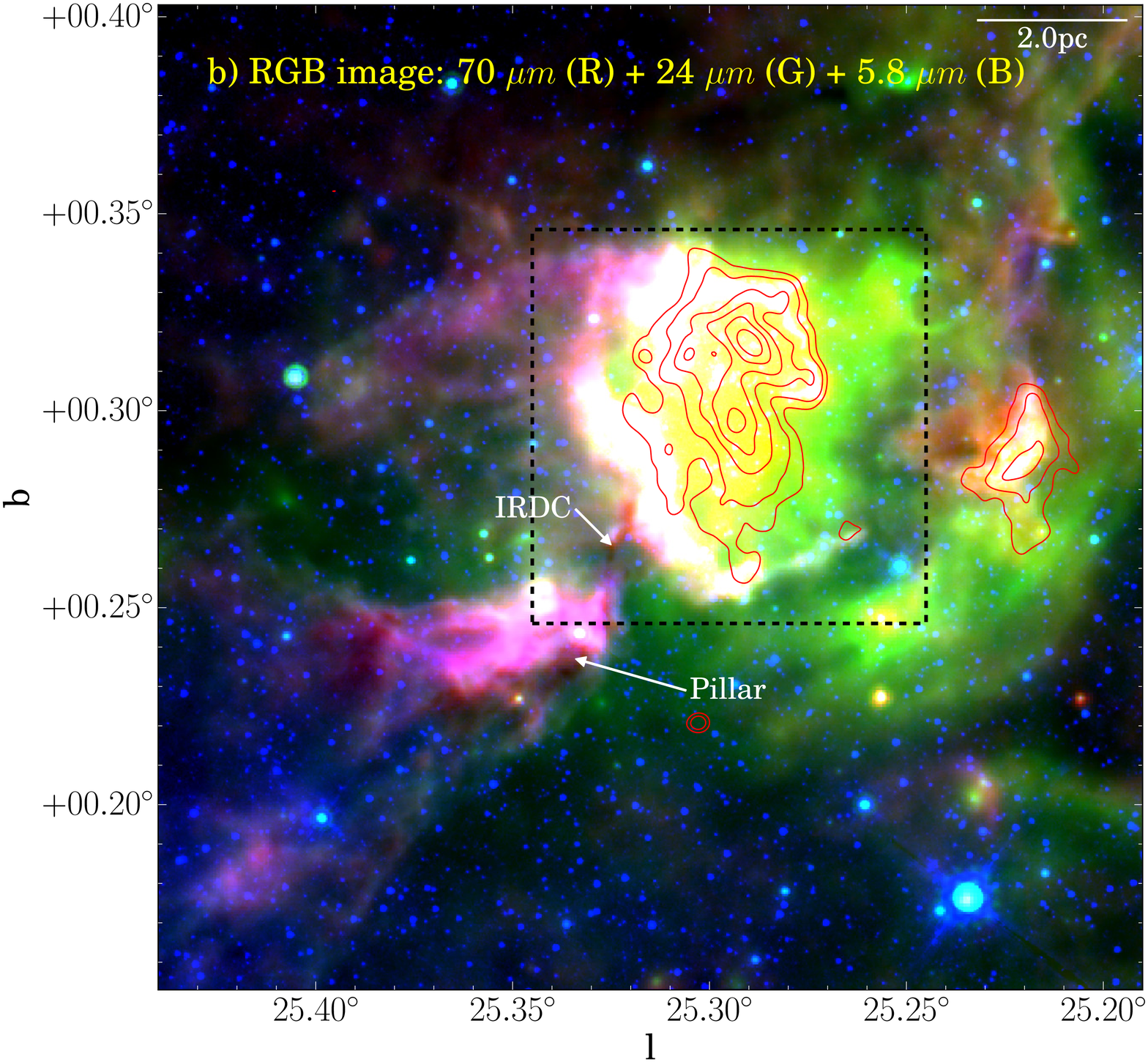}
\caption{\scriptsize A MIR and FIR view of the region around the bubble N37 (selected area $\sim$15$' \times$15$'$; centered at $l=$ 25$^\circ$.315, $b=$ 0$^\circ$.278). (a) The position
 of the bubble \citep[from][]{churchwell06} is highlighted by a red dashed ellipse in the {\it Spitzer} 8 $\mu m$ image. The star a805 \citep[O7II spectral type;][]{marco11} is highlighted in the figure.
 Position of the IRAS 18335$-$0646 is also marked. (b) Three-color composite image (red: {\it Herschel} 70 $\mu m$; green: MIPSGAL 24 $\mu m$; blue: GLIMPSE 5.8 $\mu m$) of the region around the
 bubble. The MAGPIS 20 cm radio continuum emission is shown by red contours with levels of 3.5$\sigma$, 4$\sigma$, 5$\sigma$, 6$\sigma$, 7$\sigma$ and  8$\sigma$ (where, 1$\sigma$=0.5 mJy/beam). The
 ionized emission is distributed inside the MIR bubble. The black dotted square represents the area of Figure~\ref{fig3}. The pillar-like structure and IRDC are also marked in both the panels. A scale
 bar corresponds to 2 pc (at a distance of 3 kpc) is also shown in both the images. }
\label{fig1}
\end{figure*}

In this work, we employed a multi-wavelength data to have a detailed understanding of the ongoing physical processes within and around the bubble. We selected a large-scale region of
15$' \times$15$'$ (centered at $l=$ 25$^\circ$.315, $b=$ 0$^\circ$.278) around the bubble N37, which also contains an IRDC and a pillar-like structure (see Figure~\ref{fig1}a). 
 Details of the new observations and the various archival data are described in the following sections.

%Figure 1
\subsection{Optical spectra}
To spectroscopically identify the ionizing sources of the bubble N37, we obtained the optical spectra of two point-like sources (V $\sim$14 mag) using Grism 7 and Grism 8 of the Hanle
 Faint Object Spectrograph and Camera (HFOSC; with slit width of 167 $\mu m$ average spectral resolution is $\sim$1000) attached to the 2m Himalayan {\it Chandra} Telescope
 (HCT)\footnote[1]{https://www.iiap.res.in/iao\_telescope}. Corresponding dark and flat frames were also obtained for dark-subtraction and flat-field corrections. The reduction of these
 spectra was performed using a semi-automated PyRAF based pipeline \citep{ninan14}.
\subsection{Archival Data}
We obtained the multi-wavelength data from the various Galactic plane surveys. In the following, we provide a brief description of these various archival data.
\subsubsection{Near-infrared Imaging Data}
NIR photometric $JHK$ magnitudes of point-like sources were collected from the United Kingdom Infrared Telescope (UKIRT) Infrared Deep Sky Survey (UKIDSS) Galactic Plane Survey
 \citep[GPS release 6.0; ][]{lawrence07} catalog. The UKIDSS observations were carried out using the Wide Field Camera \citep[WFCAM;][]{casali07} attached to the 3.8m UKIRT telescope.
 Spatial resolution of the UKIDSS images is $\sim$0$\farcs$8. Only good photometric magnitudes of point sources in the selected region were obtained following the conditions given
 in \citet{lucas08} and \citet{dewangan15}. Several bright sources were saturated in the UKIDSS frames. Hence, the UKIDSS sources having magnitudes brighter than J = 13.25,
 H = 12.75 and K = 12.0 mag were replaced by the Two Micron All Sky Survey \citep[2MASS;][]{skrutskie06} values.
\subsubsection {Near-infrared narrow band image}
We retrieved the H$_{2}$ (1$-$0) S(1) 2.122 $\mu$m continuum-subtracted image from the UKIRT Wide-field Infrared Survey for H$_2$ \citep[UWISH2;][]{froebrich11} archive. These observations
 were carried out using the WFCAM \citep{casali07} on the UKIRT.
\subsubsection{Near-infrared Polarization Data}
The $H$-band linear polarization data for point sources (resolution $\sim$1$\farcs$5) are also used in this study. The polarization observations were performed using the 1.8m Perkins telescope
 operated by the Boston University and the corresponding data are available in the Galactic Plane Infrared Polarization Survey \citep[GPIPS;][]{clemens12} archive. In our analysis, we only
 considered sources having good polarization measurements with P/$\sigma_P \geq$ 2.5 (where P is the degree of polarization and $\sigma_P$ is the corresponding uncertainty) and Usage Flag (UF) of 1.
\subsubsection{Mid-infrared Data}
We retrieved the 3.6, 4.5, 5.8 and 8.0 $\mu m$ images and photometric magnitudes of point sources from the {\it Spitzer}-Galactic Legacy Infrared Mid-Plane Survey Extraordinaire \citep[GLIMPSE;][]{benjamin03}
 survey (spatial resolution $\sim$2$\arcsec$). The  photometric magnitudes were obtained from the GLIMPSE-I Spring '07 highly reliable catalog. In addition to the GLIMPSE data, the Multiband Infrared
 Photometer for Spitzer (MIPS) Inner Galactic Plane Survey \citep[MIPSGAL;][]{carey05} images at 24 $\mu$m (resolution $\sim$6$\arcsec$) and the magnitudes of point sources at 24 $\mu m$ \citep{gutermuth15}
 are also used in the analysis. Some sources, which are well detected in the 24 $\mu m$ image, do not have photometric magnitudes in the MIPSGAL 24 $\mu m$ catalog of \citet{gutermuth15}. Hence,
 we separately performed the photometric reduction of 24 $\mu m$ image of the N37 region. A detailed procedure of this photometric reduction can be found in \citet{dewangan12}.
\subsubsection{Far-infrared and millimeter data}
In order to construct the dust temperature and column density maps of the N37 region, we utilized level2\_5 processed {\it Herschel} 70--500 $\mu m$ images. The beam sizes of images at
 70, 160, 250, 350, and 500 $\mu m$ are 5$\farcs$8, 12$\arcsec$, 18$\arcsec$, 25$\arcsec$, and 37$\arcsec$ \citep{poglitsch10,griffin10}, respectively.

We have also obtained the APEX Telescope Large Area Survey of Galaxy \citep[ATLASGAL;][]{schuller09} 870 $\mu m$ continuum image (beam $\sim$19$\farcs$2) and the Bolocam 1.1 mm \citep{aguirre11}
 image (beam $\sim$33$''$) of the region around the bubble.
\subsubsection {Molecular line data}
The $^{13}$CO (J=1--0) line data were retrieved from the Galactic Ring Survey \citep[GRS;][]{jackson06}. The GRS data have a velocity resolution of 0.21~km\,s$^{-1}$, an angular resolution 
 of 45$\arcsec$ with 22$\arcsec$ sampling, a main beam efficiency ($\eta_{\rm mb}$) of $\sim$0.48, a velocity coverage of $-$5 to 135~km~s$^{-1}$, and a typical rms sensitivity (1$\sigma$)
 of $\approx0.13$~K.
\subsubsection{Radio continuum data}
The Very Large Array (VLA) 20 cm radio continuum map (beam size $\sim$6$\farcs$2$\times$5$\farcs$4) of the N37 region was obtained from the Multi-Array Galactic Plane Imaging Survey archive
 \citep[MAGPIS;][]{helfand06}. 
%
%Figure 2
\begin{figure*}
\epsscale{1.0}
\plotone{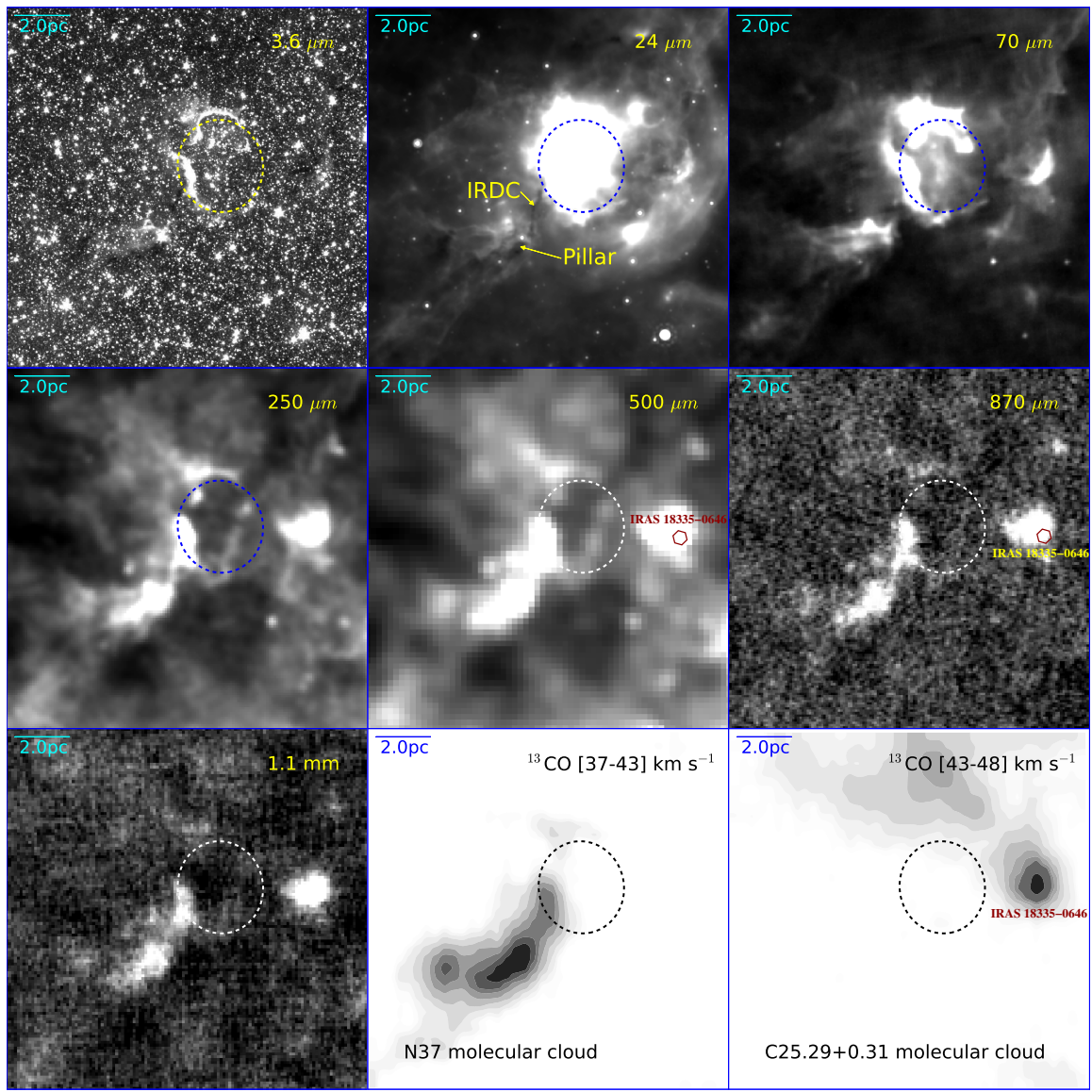}
\caption{\scriptsize Distribution of the warm and cold dust emission towards the N37 region. The images were obtained from the {\it Spitzer} (3.6--24 $\mu m$), {\it Herschel} (70--500 $\mu m$), ATLASGAL
 (870 $\mu m$), and BOLOCAM (1.1 mm) survey archives. Last two bottom panels show solid contour maps of the integrated $^{13}$CO emission in the velocity ranges of 37--43 and 43--48 km s$^{-1}$, respectively.
 The remaining symbols are similar to those shown in Figure~\ref{fig1}.}
\label{fig2}
\end{figure*}
\section{Morphology of the region}
\label{sec:morphology}
A detailed understanding of the ongoing physical processes in a given star-forming region requires a thorough and careful multi-wavelength investigation of the region. In a star-forming
 region, the spatial distribution of the ionized, dust, and molecular emission allows us to identify the H {\sc ii} regions and cold embedded condensations, which further help us to infer the
 physical conditions of the region. A multi-wavelength picture of the region around the bubble is presented in Figures~\ref{fig1} and~\ref{fig2}. In Figure~\ref{fig1}a, on a larger scale,
 the 8.0 $\mu m$ image shows a pillar-like structure, an IRDC, an IRAS source (IRAS 18335$-$0646), and the MIR bubble N37. The broken or incomplete ring morphology of N37 bubble is clearly
 seen in the image, as previously reported by \citet{churchwell06}. Figure \ref{fig1}b shows the spatial distribution of the warm dust towards the N37 region (RGB map: 70 $\mu m$ in red;
 24 $\mu m$ in green; 5.8 $\mu m$ in blue). The MAGPIS 20 cm radio continuum emission is also overlaid on the RGB map, which depicts the distribution of the ionized emission. The periphery
 of the bubble is dominated by the 5.8 $\mu m$ emission and encloses the warm dust as well as the ionized gas. In general, the polycyclic aromatic hydrocarbon (PAH) features are seen at
 3.3, 6.2, 7.7, and 8.6 $\mu$m and trace a photodissociation region (PDR) surrounding the ionized gas. Hence, the emission seen in the 5.8 and 8.0 $\mu$m images might be tracing a PDR
 towards the N37 bubble.

A longer wavelength view (250--1100 $\mu$m) of the region is presented in Figure~\ref{fig2}. The images at 3.6--70 $\mu$m are also shown for comparison with the submillimeter and millimeter
wavelength images. The emission at 250--1100 $\mu m$ traces cold dust components (see Section~\ref{colmn_tempmap} for quantitative estimates). The cold
 dust emission is mainly seen towards the pillar-like structure and the IRAS 18335$-$0646. Note that the ionized emission is also detected towards the IRAS 18335$-$0646 (see Figure~\ref{fig1}b).
 We utilized the GRS $^{13}$CO (J=1--0) line data to infer the physical association of different subregions seen in our selected region around the bubble N37. Based on the velocity information
 of $^{13}$CO data, we find that there are two molecular clouds present in our selected region. The molecular cloud associated with the bubble (i.e. N37 molecular cloud) is traced in the velocity
 range of 37--43 km s$^{-1}$. However, the molecular cloud associated with the IRAS 18335$-$0646 \citep[also referred as C25.29+0.31 in][]{anderson09} is traced in the velocity range from 43 to 48
 km s$^{-1}$. In the last two bottom panels, we show the velocity integrated $^{13}$CO maps of the two clouds seen in our selected region around the bubble. The integrated $^{13}$CO emission
 map reveals an elongated morphology of the N37 molecular cloud (i.e. velocity range $\sim$37--43 km s$^{-1}$), which hosts the pillar-like structure, an IRDC, and the bubble N37. 

On the other hand, the integrated $^{13}$CO map of the C25.29+0.31 cloud traces a large condensation associated with the IRAS 18335$-$0646, as seen in the longer wavelength continuum images
 (see Figure~\ref{fig2}). \citet{wienen12} also reported the NH$_{3}$ line parameters such as NH$_{3}$(1,1) radial velocity $\sim$46.13 km s$^{-1}$ and kinematic temperature (T$_{kin}$)
 $\sim$21.76 K toward the ATLASGAL condensation associated with the IRAS 18335$-$0646. These results suggest that the condensation associated with the IRAS 18335$-$0646 is not physically
 linked with the bubble N37.

A more detailed analysis of these two molecular clouds (i.e. N37 molecular cloud and C25.29+0.31) is discussed in the Section~\ref{sec:CO}.
\section{Results}
\label{sec:result}
In this section, we present the outcomes of our multi-wavelength analysis in the following way. First, we present the results related to the identification of ionizing source(s) of the region,
 and then the origin of the N37 bubble. Next, we present the column density and the temperature maps of the region to identify the cold condensations. We have also identified young stellar
 objects (YSOs) towards the region, and construct the surface density map of these YSOs to study their spatial distribution. Finally, we examine the NIR polarization data and the $^{13}$CO
 molecular line data to examine the large scale magnetic field morphology and the kinematics of CO gas, respectively.
%
%Figure 3
\begin{figure}
\epsscale{1.1}
\plotone{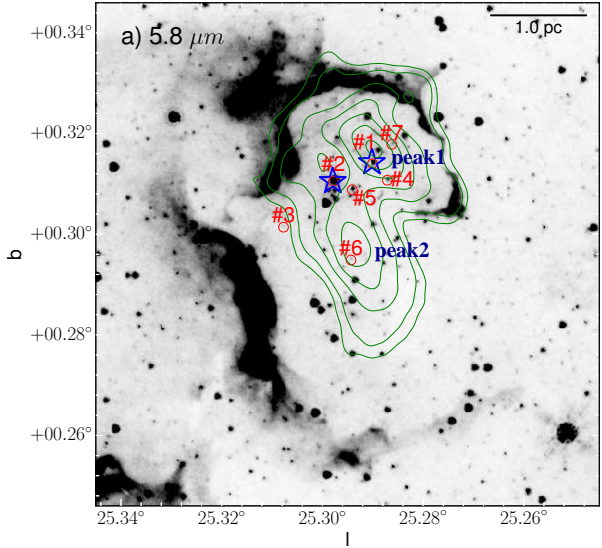}
\caption{\scriptsize Photometrically identified OB stars are marked by open circles and are labeled by numbers (as per their respective serial numbers in Table~\ref{table1}) on the {\it Spitzer}-IRAC
 5.8 $\mu m$ image. The MAGPIS 20 cm radio contours are also overlaid on the image. Asterisks represent the sources for which optical spectroscopic observations were performed (see Figure~\ref{fig4}).}
\label{fig3}
\end{figure}
\subsection{Identification of ionizing candidates} 
\label{PhotometricIonizing}
We have seen that the ionized emission is enclosed within the N37 bubble (see Figure~\ref{fig1}b) and two prominent peaks (i.e. peak1 and peak2 shown in Figure~\ref{fig3}) are seen in the radio
 continuum map. To search for possible OB type candidates located within the N37 bubble, we performed a photometric method to identify the probable ionizing candidates of the region,
 following a similar procedure outlined in \citet{dewangan13}. The analysis was carried out using the NIR and MIR photometric magnitudes of point sources from the UKIDSS and GLIMPSE catalogs,
 respectively. We only considered the sources located near the radio emission peaks and detected at least in five photometric bands among UKIDSS JHK and {\it Spitzer}-IRAC 3.6, 4.5 and
 5.8 $\mu m$ bands. Following this condition, a total of seven sources were identified, and these are marked and labeled in Figure~\ref{fig3}. The extinction to these sources were estimated
 assuming intrinsic colors of (J-H)$_0$ and (H-K)$_0$ for O- and B-stars from \citet{martins06} and \citet{pecaut13}, respectively,
 and using the extinction law ($A_J / A_V$ = 0.284, $A_H / A_V$ = 0.174, $A_K / A_V$ = 0.114) from \citet{indebetouw05}. The absolute JHK magnitudes of all these sources were calculated
 assuming a distance of 3.0 kpc and were compared with those listed in \citet[][for O stars]{martins06}, and \citet[][for B stars]{pecaut13}. We found that two O-type and five B-type stars
 are located near the radio peaks within the N37 bubble. All these sources with their photometric magnitudes and derived spectral types are listed in Table~\ref{table1}. Note that a single
 distance is assumed for all the sources, and the spectral types are also estimated without considering the photometric uncertainties of intrinsic colors and observed magnitudes of these
 sources. Spectroscopic observations will be helpful to further confirm the spectral type of these sources (see Section~\ref{OpticalSpec}).

%Table 1
\begin{deluxetable*}{llccccccccccccc}
%\rotate
\tablewidth{0pt}
\tabletypesize{\scriptsize} 
\setlength{\tabcolsep}{0.065in}
\tablecaption{Details of the photometrically identified OB stars toward the N37 region.\label{table1}}
\tablehead{
\colhead{Sr.}&\colhead{RA (J2000)}&\colhead{Dec (J2000)}&\colhead{J}   &\colhead{H}   &\colhead{K}    &\colhead{[3.6]}&\colhead{[4.5]}&\colhead{[5.8]}&\colhead{[8.0]}&\colhead{$A_V$}& \colhead{$M_J$}& \colhead{$M_H$}& \colhead{$M_K$}& \colhead{Sp. Type} \\  
\colhead{No.}&\colhead{(hh:mm:ss)}&\colhead{(dd:mm:ss)}&\colhead{(mag)}&\colhead{(mag}&\colhead{(mag)}&\colhead{(mag)}&\colhead{(mag)}&\colhead{(mag)}&\colhead{(mag)}&\colhead{(mag)}& \colhead{(mag)}& \colhead{(mag)}& \colhead{(mag)}&             }
\startdata
1  & 18:36:18.6 & -06:39:09 & 10.17 &  9.62 &  9.34 & 10.17 &  9.62 &  9.34 &  9.24 &  6.11 & -3.95 & -3.84 & -3.74 & O8V$^a$  \\
2  & 18:36:20.2 & -06:38:50 & 10.41 &  9.80 &  9.38 & 10.41 &  9.80 &  9.38 &  8.15 &  7.49 & -4.10 & -3.90 & -3.86 & O7V-O8V$^b$\\
3  & 18:36:23.3 & -06:38:34 & 15.23 & 13.44 & 12.55 & 15.51 & 13.53 & 12.70 & 11.91 & 16.20 & -1.76 & -1.80 & -1.67 & B2V      \\
4  & 18:36:19.0 & -06:39:24 & 13.61 & 12.64 & 11.92 & 13.76 & 12.62 & 11.93 & 11.06 & 11.05 & -1.92 & -1.70 & -1.73 & B2V      \\
5  & 18:36:20.2 & -06:39:05 & 14.12 & 12.88 & 12.14 & 13.73 & 12.42 & 11.73 &   --  & 12.45 & -1.80 & -1.70 & -1.66 & B2V      \\
6  & 18:36:23.2 & -06:39:27 & 14.52 & 12.98 & 12.20 & 14.42 & 12.71 & 11.94 & 11.38 & 14.21 & -1.90 & -1.90 & -1.81 & B2V      \\
7  & 18:36:17.4 & -06:39:15 & 15.29 & 13.18 & 12.08 & 15.20 & 13.04 & 11.94 & 11.18 & 19.77 & -2.71 & -2.69 & -2.56 & B1V      
\enddata
\tablenotetext{a}{Spectroscopically identified O9V}
\tablenotetext{b}{Spectroscopically identified B0V}
\end{deluxetable*}
\subsection{Spectroscopy of ionizing candidates} 
\label{OpticalSpec}
\begin{figure*} 
 \epsscale{1}
\plotone{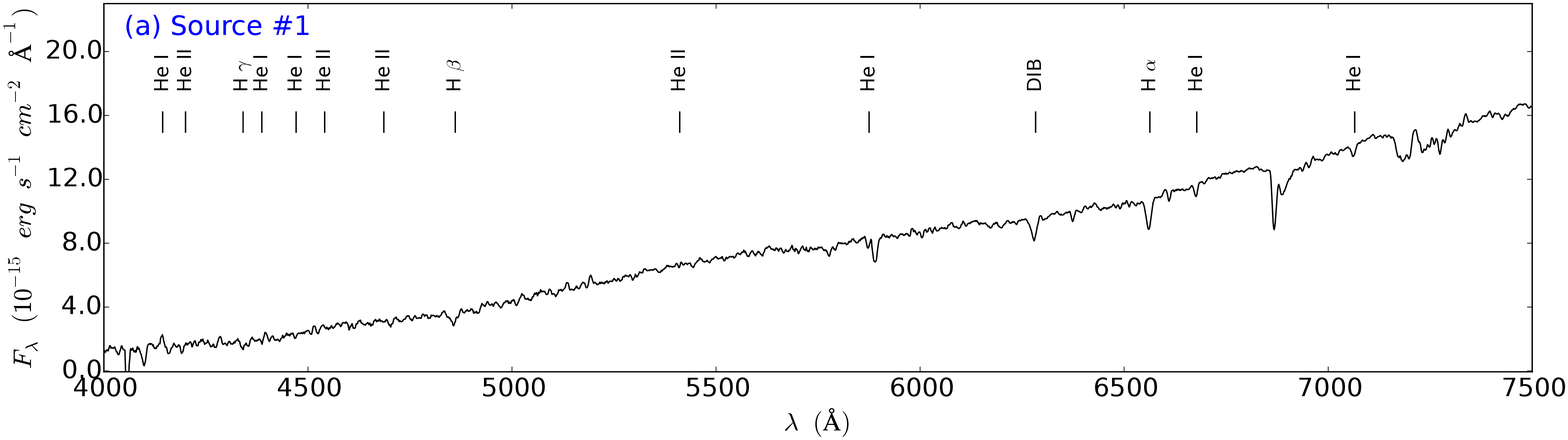}
\epsscale{1}
\plotone{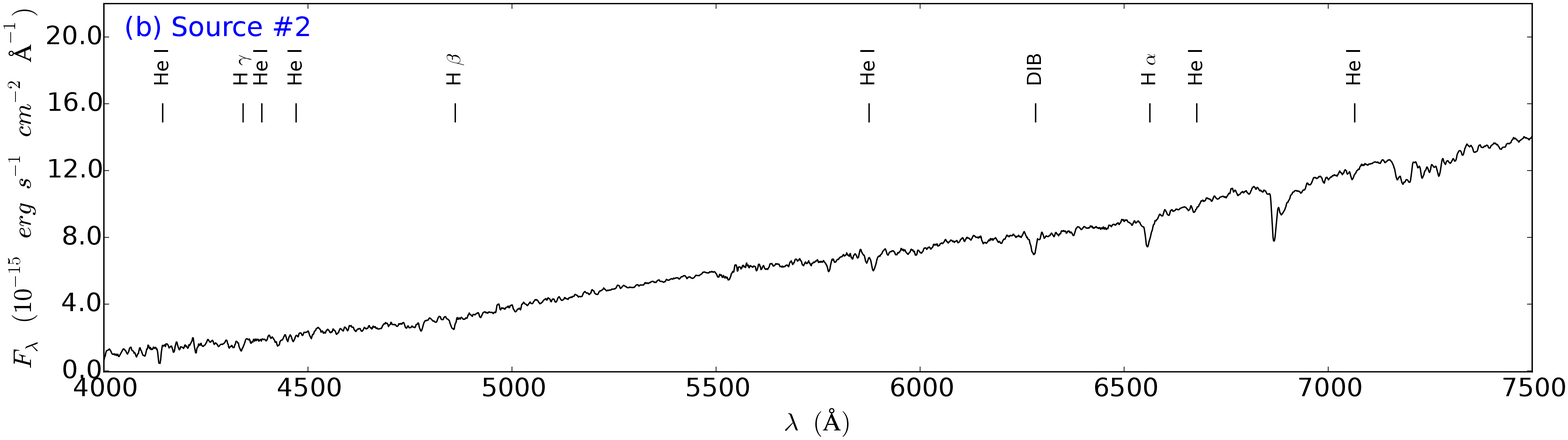}
\caption{Optical spectra of sources \#1 (a) and \#2 (b) (as marked in Figure \ref{fig3}) observed using the HFOSC mounted on the 2-m HCT. Several prominent hydrogen and helium lines are detected in both the spectra.}
\label{fig4}
\end{figure*}

We have carried out optical spectroscopic observations (4000-7500 \AA) of two brightest photometrically identified OB stars (see asterisks in Figure~\ref{fig3}). The remaining five
 sources are beyond the limit of optical spectroscopic capability of the HCT (V-limit $\simeq$18-mag for spectrum having signal-to-noise ratio of about 20 with 30 min exposure). The observed
 spectra of these two sources are shown in Figure \ref{fig4}. One of these sources (source \#1) is situated near the peak1 ($\sim$10$''$) of the 20 cm radio emission, and the other one
 (\#2) is located $\sim$36$''$ away from the radio peak1.
 
Several hydrogen and helium lines are found in both the spectra (Figures~\ref{fig4}a and~\ref{fig4}b). The presence of hydrogen lines is generally seen in early type sources (O--A spectral type),
 however, the existence of He {\sc i-ii} lines is not found in A stars \citep{walborn90}. Note that the ionization of helium requires a high temperature generally seen in
 O-type sources. To confirm the spectral types of these sources, we further compared our observed spectra with the available OB stars' spectra \citep{walborn90,pickles98}.
 Generally, spectral lines in the first part of the optical spectrum (4000-5500 \AA) are used to  determine the spectral type of any source. However, the signal-to-noise ratio of
 the first part of both the spectra is not very good, possibly because of large visual extinction toward the region. A visual comparison of the observed spectra with the available
 library spectra reveals that the source \#1 is likely to be a O9V star, while the other source (\#2) is an B0V candidate. It can be seen in Section~\ref{PhotometricIonizing} that the
 second source was photometrically identified as O7-8V star possibly because of the assumed distance of 3.0 kpc in the calculation which might not be true (see next paragraph).

%Figure 4
%
We have also estimated spectro-photometric distances to these two sources. The optical BV-band magnitudes of both the sources were obtained from the American Association of Variable Stars
 Observers (AAVSO) Photometric All-Sky Survey (APASS) catalog (DR9) and the NIR $K$-band magnitudes were collected from the 2MASS catalog. To estimate the distance to the O9V source, we
 first obtained the intrinsic color of an O9V star ($(V-K)_o \sim$--0.79) from \citet{wegner94}. Furthermore, the color excess $E(B-V)$ was estimated using the relation $E(V-K)$/$E(B-V)\sim$
 2.826 given in \citet{wegner94}. Though it is debatable whether the value of $R_V$ ($A_V$/$E(B-V)$) is 3.1 all over the Galaxy or is it substantially different in Galactic star-forming
 regions \citep[see][and references therein]{pandey03}, we considered the mean $R_V$ of 3.1 itself for the reddening correction. Accordingly, the visual extinction of the source was found
 to be 6.3 mag. The absolute and apparent V-band magnitudes of both the sources were obtained from \citet{lang99} and the APASS catalog, respectively. With an absolute V-magnitude of -4.5 and
 apparent V-magnitude of 14.37, we estimated the distance to the O9V star of 3.1 kpc. Following the similar procedure for the other source (B0V) with the absolute and apparent
 magnitudes of -4.0 and 14.55, respectively, the visual extinction ($A_V$) was estimated to be 6.4 mag, and the corresponding distance to the source is 2.7 kpc. Note that large errors
 (at least 20\%) could be associated with these distance estimates due to photometric uncertainties, and the general extinction law used in the estimation. However, similar distances of
 these sources and the bubble suggest that they are physically associated with the N37 bubble.
\subsection{Radio continuum emission and the dynamical age}
\label{sec:radio}
The integrated radio continuum flux is used as a tool to determine the spectral type of the source responsible to develop the H {\sc ii} region. The presence of an H {\sc ii} region is
 traced in the MAGPIS 20 cm map (see Figure \ref{fig1}b). The Lyman continuum flux (photons s$^{-1}$) required for the observed radio continuum emission is estimated following the
 equation given in \citet{moran83}:
\begin{equation}
  S_{Lyc} = 8 \times 10^{43}\left(\frac{S_\nu}{mJy}\right)\left(\frac{T_e}{10^4 K}\right)^{-0.45}\left(\frac{D}{kpc}\right)^2 \left(\frac{\nu}{GHz}\right)^{0.1}
\label{lyman_flux}
\end{equation}
where $\nu$ is the frequency of observations, $S_\nu$ is the total observed flux density, $T_e$ is the electron temperature, and $D$ is the distance to the source. Here, the region is
 assumed to be homogeneous and spherically symmetric, and a single main-sequence star is responsible for the observed free-free emission. The flux density ($S_\nu$) and the size of the
 H {\sc ii} region are determined using the {\sc jmfit} task of the Astronomical Image Processing Software (AIPS). Typical value of the electron temperature, $T_e \sim$ 10000 K for a
 classical H {\sc ii} region \citep{stahler05} is adopted in the calculation. The spectral type of the powering source is finally estimated by comparing the observed Lyman continuum
 flux with the theoretical value for solar abundance given in \citet{smith02}. 
 
In the MAGPIS 20 cm map, two radio peaks are clearly evident within the bubble (peak1 and peak2; see Figure~\ref{fig3}), and we estimated the spectral type of the possible ionizing source
 for both the peaks separately. The Lyman continuum flux for the radio peak1 (S$_\nu \sim$ 1.18 Jy; S$_{Lyc} \sim$ 10$^{47.95}$ photons sec$^{-1}$) corresponds to an ionizing source having
 spectral type of O9V, while the ionizing source corresponding to the radio peak2 (S$_\nu \sim$ 0.56 Jy; S$_{Lyc} \sim$ 10$^{47.62}$ photons sec$^{-1}$) is a B0V star.

As mentioned before (see Sections~\ref{PhotometricIonizing} and \ref{OpticalSpec}), using the spectroscopy and photometry, we identified three OB stars (O9V, B1V, and B2V) that are located
 near the radio peak1 (see Figure~\ref{fig3}). However, the Lyman continuum flux \citep{smith02} expected together from solar abundant B1V and B2V stars is about an order less compared to
 the O9V star, and therefore, the total flux is mainly dominated by the O9V star. Hence, it seems that an O9V star located at a distance of $\sim$10$''$ from the radio peak1 is the primary
 ionizing source of the region. The spectral type of the ionizing source determined from the radio analysis is consistent with our spectroscopic results. Toward the radio peak2, we identified
 a B2V star using the photometric analysis, however the estimation from the radio continuum flux shows the ionizing source to be a B0V star. The spectral type corresponding to the peak2
 estimated using two methods is showing inconsistency because the photometric determinations of spectral types may vary substantially depending on the distance to the source and the photometric accuracy.

We have also determined the dynamical age of the H {\sc ii} region using the MAGPIS 20 cm data. A massive source ionizes the surrounding gas, and develop an H {\sc ii} region. The ionization
 front of the H {\sc ii} region expands until an equilibrium is achieved between the rate of ionization and recombination. Theoretical radius of the H {\sc ii} region \citep[i.e., Str\"omgren
 radius;][]{stromgren39} for a uniform density and temperature, can be written as:
\begin{equation}
R_S = \left(\frac{3S_{Lyc}}{4\pi n_0^2\beta_2}\right)^{1/3}
\end{equation}
where $n_0$ is the initial ambient density, and $\beta_2$ is the recombination coefficient. For a temperature of 10,000 K, the value of $\beta_2$ is 2.60$\times$10$^{-13}$ cm$^3$ s$^{-1}$ \citep{stahler05}. 

A shock front is generated because of the large temperature and pressure gradient between the ionized gas and the surrounding cold material, and the shock front is further propagated into the
 surroundings. The corresponding radius of the ionized region at any given time can be written as \citep{spitzer78}:
\begin{equation}
    R(t) = R_S \left(1 + \frac{7c_{II} t_{dyn}}{4R_S}\right)^{4/7}
\end{equation}
where the speed of sound in an H {\sc ii} region (c$_{II}$) is 11$\times$10$^5$ cm s$^{-1}$ \citep{stahler05} and $t_{dyn}$ is the dynamical age of the H {\sc ii} region. The size of the
 H {\sc ii} region, i.e., R(t), was estimated to be 1.4 pc by using the {\sc jmfit} task of the AIPS. Note that the calculated dynamical age can vary substantially depending on the initial value
 of the ambient density. Therefore, we estimated the Str\"{o}mgren radius and the corresponding dynamical age for a range of ambient density starting from 1000 to
 10000 cm$^{-3}$ \citep[e.g. classical to ultra-compact H {\sc ii} regions;][]{kurtz02}. For the corresponding densities, the dynamical age varies from 0.21--0.71 Myr. However, the Str\"{o}mgren
 radius and dynamical age were calculated by assuming the region as homogeneous and spherically symmetric. Hence, the dynamical age of the H {\sc ii} region should be considered as a representative
 value (see Section~\ref{sec:discussion} for more discussion).
\subsection{Origin of the bubble}
\label{sec:ratt}
\begin{figure} 
 \epsscale{1.1}
\plotone{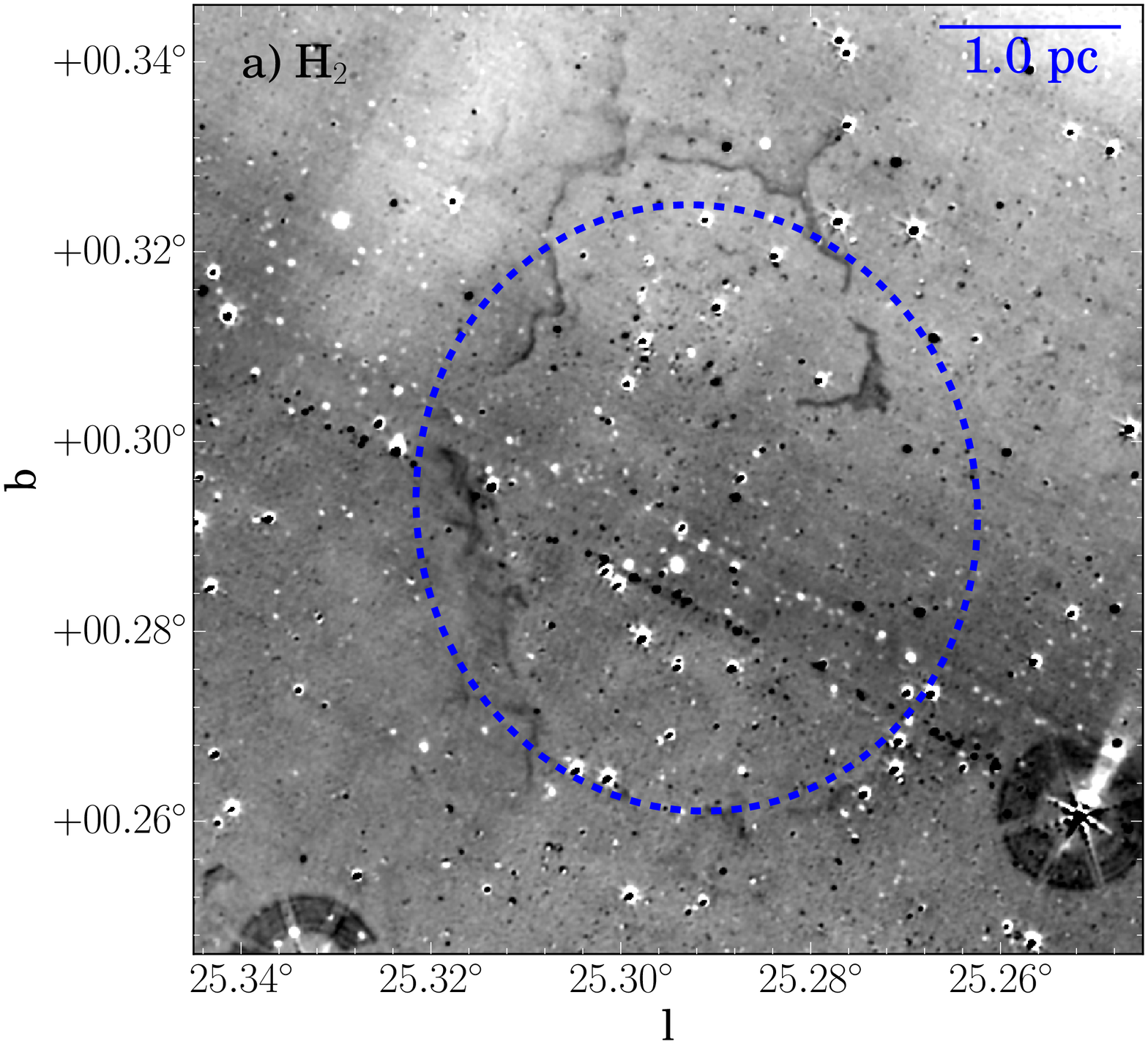}
 \epsscale{1.1}
\plotone{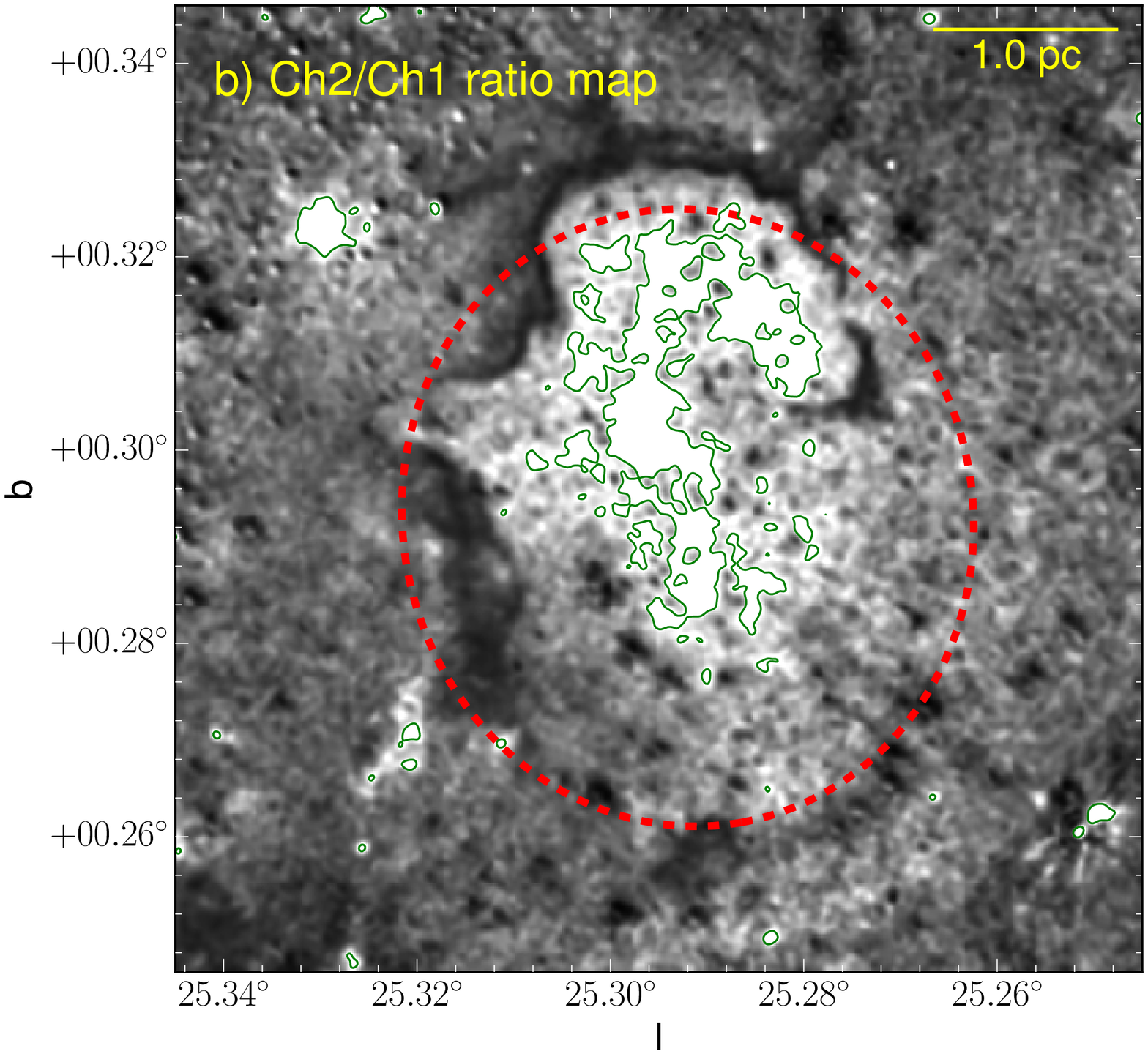}
\caption{\scriptsize (a) Inverted continuum-subtracted H$_2$-band ($\nu$ = 1--0 S(1) at 2.122 $\mu m$) image of the N37 region. (b) {\it Spitzer}-IRAC Ch2/Ch1 ratio map of the N37 region. In the ratio map, the
 bright emission region is traced by contours. The interior of the bubble is traced with ratio greater than 1.1, while the periphery of the bubble is traced with ratio less than 0.8.}
\label{fig5}
\end{figure}

In Figure~\ref{fig5}a, we present the continuum-subtracted H$_2$ emission (2.122 $\mu m$) map toward the bubble N37, which traces the edges of the bubble. The H$_2$ features have similar
 morphology as seen in the {\it Spitzer}-GLIMPSE images. The H$_2$ emission in a given star-forming region is originated in the shocked region developed at the interface of the ionized and
 cold matter. From the distribution of H$_{2}$ emission, 8.0 $\mu$m emission and the ionized emission (see Figure~\ref{fig2}), it is evident that the emission seen in the narrow H$_{2}$-band
 is tracing the PDR towards the N37 region. Possibly, the H$_{2}$ emission is originated in the shocked region developed due to the expansion of the ionized gas. 

Ratio maps of {\it Spitzer}-IRAC images have ability to provide the information about the interaction of massive star(s) with its surrounding environment \citep{povich07, dewangan12, dewangan13}. Note that the
 {\it Spitzer}-IRAC bands contain several prominent characteristic atomic and molecular lines. For example, IRAC Ch1 contains a PAH feature at 3.3 $\mu m$ as well as a prominent molecular
 hydrogen line at 3.234 $\mu$m ($\nu$ = 1--0 $ O$(5)). IRAC Ch2 also contains a molecular hydrogen emission line at 4.693 $\mu m$ ($\nu$ = 0--0 $S$(9)) generally excited by
 outflow shocks, and a hydrogen recombination line Br$\alpha$ (4.05 $\mu$m). Figure~\ref{fig5}b shows the IRAC Ch2/Ch1 ratio map of the N37 region, which reveals the bright emission region
 surrounded by the dark features. In general, the dark regions in the 4.5 $\mu$m/3.6 $\mu$m ratio map traces the excess 3.6 $\mu$m emission, while the bright emission region suggests the
 domination of 4.5 $\mu$m emission. The bright emission region in the ratio map is very well correlated with the radio continuum emission. Therefore, it seems that this bright emission region
 probably traces the Br$\alpha$ feature originated by the photoionized gas. The dark features in the ratio map are also well correlated with the 2.122 $\mu$m H$_{2}$ emission, indicating that
 the ratio map probably traces the H$_{2}$ features (see Figures~\ref{fig5}a and~\ref{fig5}b). However, it must be noted that the Ch1 also contains 3.3 $\mu m$ PAH emission feature which
 may also contribute to the dark features seen in the ratio map. Overall, we found that the IRAC ratio map and the continuum-subtracted H$_{2}$ image trace PDR around the H\,{\sc ii} region.

%Figure 5
%
Massive stars can influence their parent molecular clouds via different feedback components - (i) pressure due to radiation (P$_{rad}$), (ii) pressure due to H{\sc ii} region (P$_{H II}$),
 and (iii) pressure due to wind (P$_{wind}$). It is important to determine the strongest pressure component of massive stars in order to have a better knowledge of the feedback mechanisms.
 Pressure due to radiation can be formulated as P$_{rad}$ = $L_{bol}/4\pi c D_s^2$, where $L_{bol}$ is the bolometric luminosity and D$_S$ is the distance from the star to the region of
 interest. Similarly, the pressures due to the H {\sc ii} region and stellar wind can be written as P$_{H II}$ = $\mu m_H C_{II}^2$ $\left(3N_{UV}/4\pi\beta_2 D_S^3\right)^{1/2}$ and P$_{wind}$ =
 $\dot{M_w} V_w/4\pi D_S^2$, respectively, where the mean molecular weight in an H {\sc ii} region, $\mu$=0.678 \citep{bisbas09}, C$_{II}$ is the sound speed in an H {\sc ii} region = 11 km s$^{-1}$,
 $\beta_2$ is a recombination coefficient = 2.6$\times$10$^{-13}$ cm$^3$ s$^{-1}$, N$_{UV}$ is the number of UV photons, $\dot{M_w}$ is the mass-loss rate of the source, $V_w$ is the terminal
 velocity of the stellar wind \citep[see][for more details about these formulas]{bressert12}. All the pressure components were estimated at a distance of $\sim$1 pc which is the nearest
 edge of the bubble from the massive O9V star.
 
The pressure exhibited by the H {\sc ii} region with N$_{UV} \sim$ 10$^{48.12}$ photons sec$^{-1}$ (i.e., total Lyman continuum for both the radio continuum peaks; see Section~\ref{sec:radio})
 is estimated to be 2.8$\times$10$^{-10}$ dyne cm$^{-2}$. For the estimation of P$_{rad}$, the bolometric luminosities of all seven OB stars within the bubble were obtained from \citet{lang99} and
 the total radiation pressure exhibited by all these OB stars is found to be 1.8$\times$10$^{-10}$ dyne cm$^{-2}$. To estimate the combined P$_{wind}$ from all the seven OB stars, the wind speed
 and the mass-loss rate for the O9V star were obtained from \citet{muijres12} (V$_w \sim$1000 km s$^{-1}$; $\dot{M_w}\sim$10$^{-7.8} M_\odot$ yr$^{-1}$). The corresponding values for B0V, B1V and
 B2V stars (V$_w \sim$1000, 700 and 700 km s$^{-1}$; $\dot{M_w}\sim$10$^{-9.3}$, 10$^{-9.4}$; 10$^{-9.7}$ M$_\odot$ yr$^{-1}$, respectively) were obtained from \citet{oskinova11}. The combined
 pressure due to the wind (P$_{wind}$) from all the seven OB stars comes out to be 5.1$\times$10$^{-12}$ dyne cm$^{-2}$. The estimation of different pressure components infers that the pressure due
 to the ionized gas (i.e, H {\sc ii} region) is the predominant component. However, there is also a substantial radiation pressure contributed together by all the OB stars. Note five out of seven of these
 OB stars not only lack of spectroscopic confirmations, but also association of them with the N37 bubble is uncertain. Therefore, the calculated values of the pressure due to the radiation and the stellar
 wind should be treated as upper limits. From the overall analysis it seems that a shock-front has been developed due to the expansion of the H {\sc ii} region which excites
 the H$_2$ emission as well as PAH emission. Our results indicate that the N37 bubble is possibly originated due to the ionizing feedback of the massive stars.
\subsection{Column density and temperature maps}
\label{colmn_tempmap}
We have constructed the column density and the temperature maps of the region using {\it Herschel} images to probe the condensations and the distribution of cold matter. A pixel-by-pixel
 modified blackbody fit was performed to the cold dust emission seen in the {\it Herschel} 160, 250, 350 and 500 $\mu m$ images. The 70 $\mu m$ image was not considered in our
 analysis because a substantial part of the 70 $\mu m$ flux comes from the warm dust. Before performing the fit, all the images were convolved to the lowest resolution of
 37$''$ (beam size of the 500 $\mu m$ image) and converted to the same flux unit (Jy pixel$^{-1}$). For better estimation of the source flux, we subtracted the corresponding background
 flux from each image \citep[see][for more detail]{mallick15}. Background flux was estimated in a relatively dark region ($l=$ 24$^\circ$.60, $b=$ 1$^\circ$.00;
 area: 10$' \times$10$'$) away from our selected target, and corresponding fluxes are -2.202, 1.328, 0.693 and 0.252 Jy pixel$^{-1}$ for 160, 250, 350 and 500 $\mu m$ images,
 respectively.
 
Finally, the pixel-by-pixel basis modified blackbody fitting was performed by using the formula \citep{battersby11,sadavoy12,nielbock12,launhardt13}:
\begin{equation}
S_\nu (\nu) - I_{bg} (\nu) = B_\nu(\nu, T_d) \Omega (1 - e^{-\tau(\nu)})
\end{equation}
where optical depth can be written as:
\begin{equation}
\tau(\nu) = \mu_{H_2} m_{H} \kappa_\nu N(H_2)
\end{equation}
Different symbols in the above equations are as follows - $S_\nu (\nu)$: observed flux density, $I_{bg}$: background flux density, $B_\nu(\nu, T_d)$: Planck's
 function, $T_d$: dust temperature, $\Omega$: solid angle subtended by a pixel, $\mu_{H_2}$: mean molecular weight, $m_H$: mass of hydrogen, $\kappa_\nu$: dust
 absorption coefficient, and $N(H_2)$: column density. Here, we used $\Omega$ = 4.612$\times$10$^{-9}$ steradian (i.e. for 14$\arcsec\times$14$\arcsec$
 area), $\mu_{H_2}$\,=\,2.8 and $\kappa_\nu$\,=\,0.1~$(\nu/1000~{\rm GHz})^{\beta}$ cm$^{2}$ g$^{-1}$, the gas-to-dust ratio of 100, and the dust spectral index
 $\beta$\,=\,2 for sources with thermal emission in the optically thick medium \citep{hildebrand83}.

\begin{figure} 
 \epsscale{1.2}
\plotone{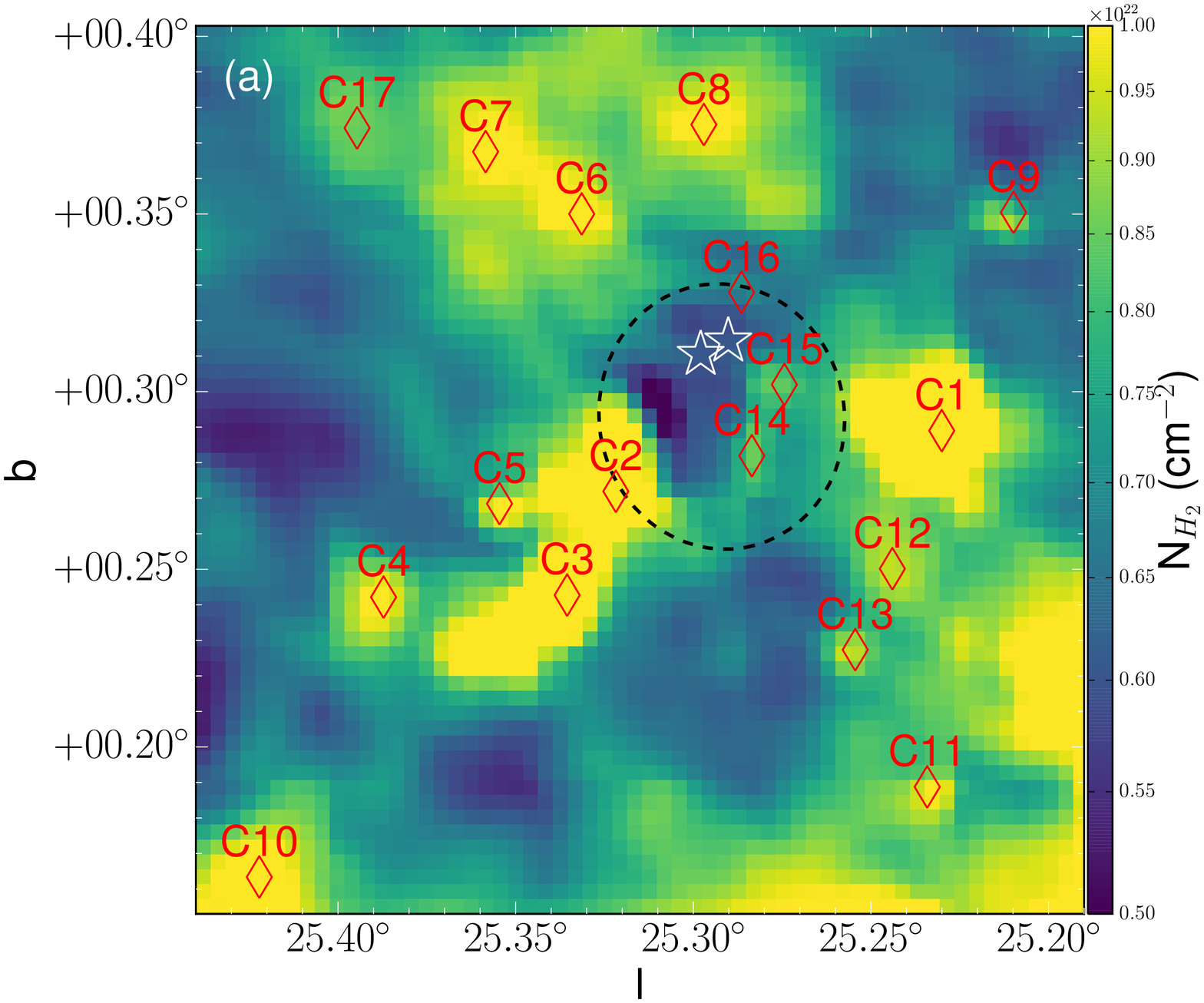}
 \epsscale{1.2}
\plotone{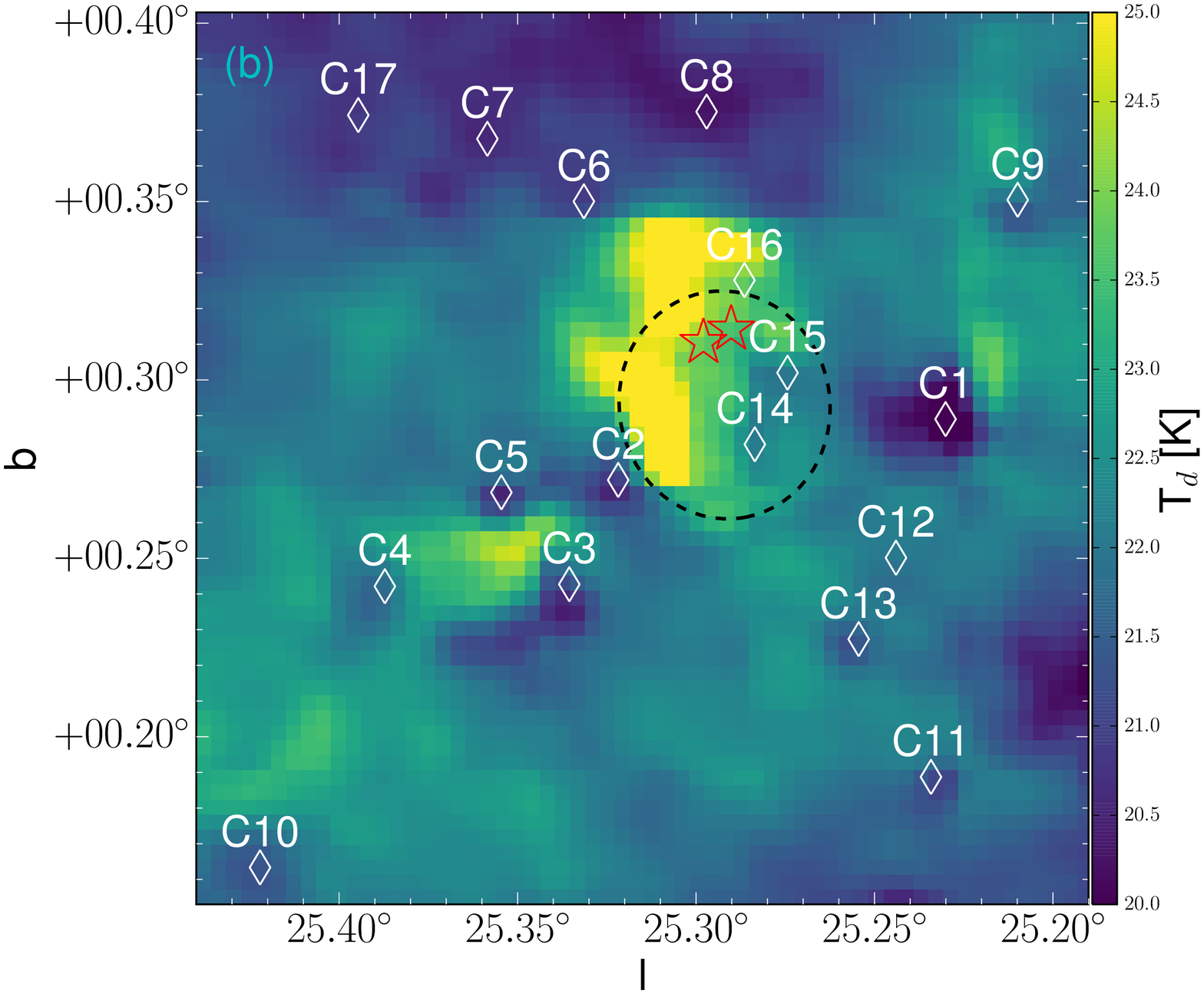}
\caption{\scriptsize {\it Herschel} (a) column density and (b) temperature maps of the selected area around the N37 bubble. Several identified cold condensations are also marked ($\diamond$) in both the maps.}
\label{fig6}
\end{figure}

The final column density and temperature maps of the 15$' \times$15$'$ area of the N37 region are shown in Figures~\ref{fig6}a and~\ref{fig6}b, respectively. Several condensations are seen
 towards the region. The `clumpfind' software \citep{williams94} has been used to identify the clumps and to measure the total column density in each clump. The mass
 of a clump is estimated using the formula \citep{mallick15}:
\begin{equation}
M_{clump} = \mu_{H_2} m_H Area_{pix} \Sigma N(H_2)
\end{equation}
where $\mu_{H_2}$\,=\,2.8, $Area_{pix}$ is the area subtended by one pixel, and $\Sigma N(H_2)$ is the total column density of the clump obtained using the `clumpfind'. A total of 17 clumps
 are identified toward the 15$' \times$15$'$ area of the N37 region (see Figure~\ref{fig6}). However, based on the integrated CO maps (see Figure~\ref{fig2}), we find only five clumps (i.e.
 C2--5 and C10, in Figure~\ref{fig6}) associated with the N37 molecular cloud, and the remaining clumps appear to be associated with the C25.29+0.31 molecular cloud. In the present work, our analysis is focused on
 the N37 molecular cloud, and hence, we do not discuss the results of the C25.29+0.31 cloud. In the N37 molecular cloud, the five associated clumps (i.e. C2--5 and C10; M$_{clump}$ from $\sim$1350--2150
 M$_{\odot}$) are having temperatures and densities in the range from $\sim$22--23~K and $\sim$7.2--9.5~$\times$~10$^{21}$ cm$^{-2}$ (corresponding A$_{V}$ $\sim$8.0--10.0 mag), respectively.
 Here, to estimate the visual extinction, we use the relation $\langle N(H_2)/ A_V \rangle = 0.94 \times 10^{21}$ molecules cm$^{-2}$ mag$^{-1}$ \citep{bohlin78}.
%
%Figure 6
\subsection{Young stellar population}
The study of YSOs in a given star-forming region allows to characterize the area of the ongoing star formation. Hence, we have carried out identification of YSOs using the NIR and MIR color-magnitude
 and color-color schemes. A more elaborative description of these schemes is given below.
%
%Figure 7
\begin{figure*}
\begin{tabular}{cc}
    \includegraphics[width=0.45\textwidth]{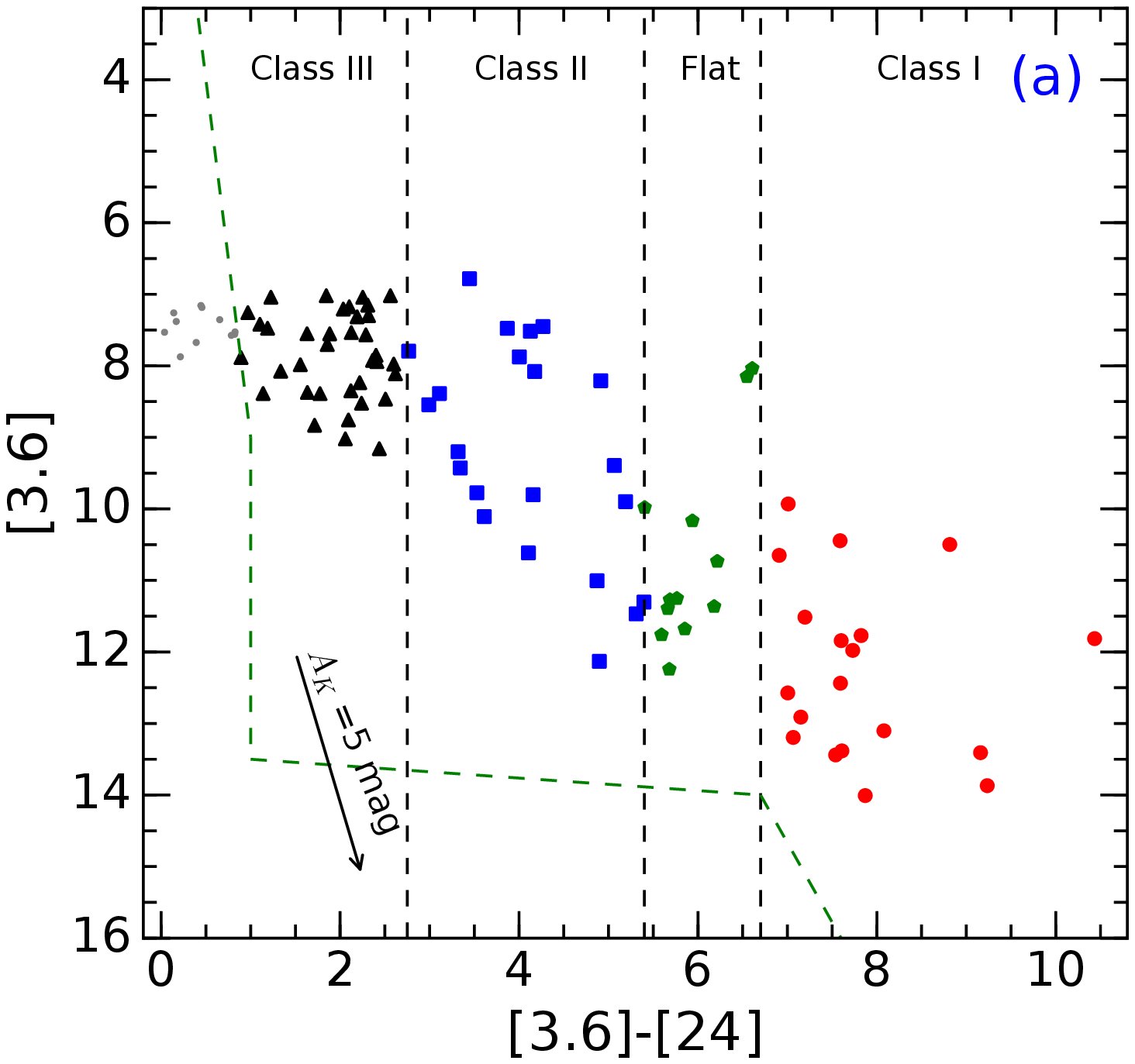} & \includegraphics[width=0.45\textwidth]{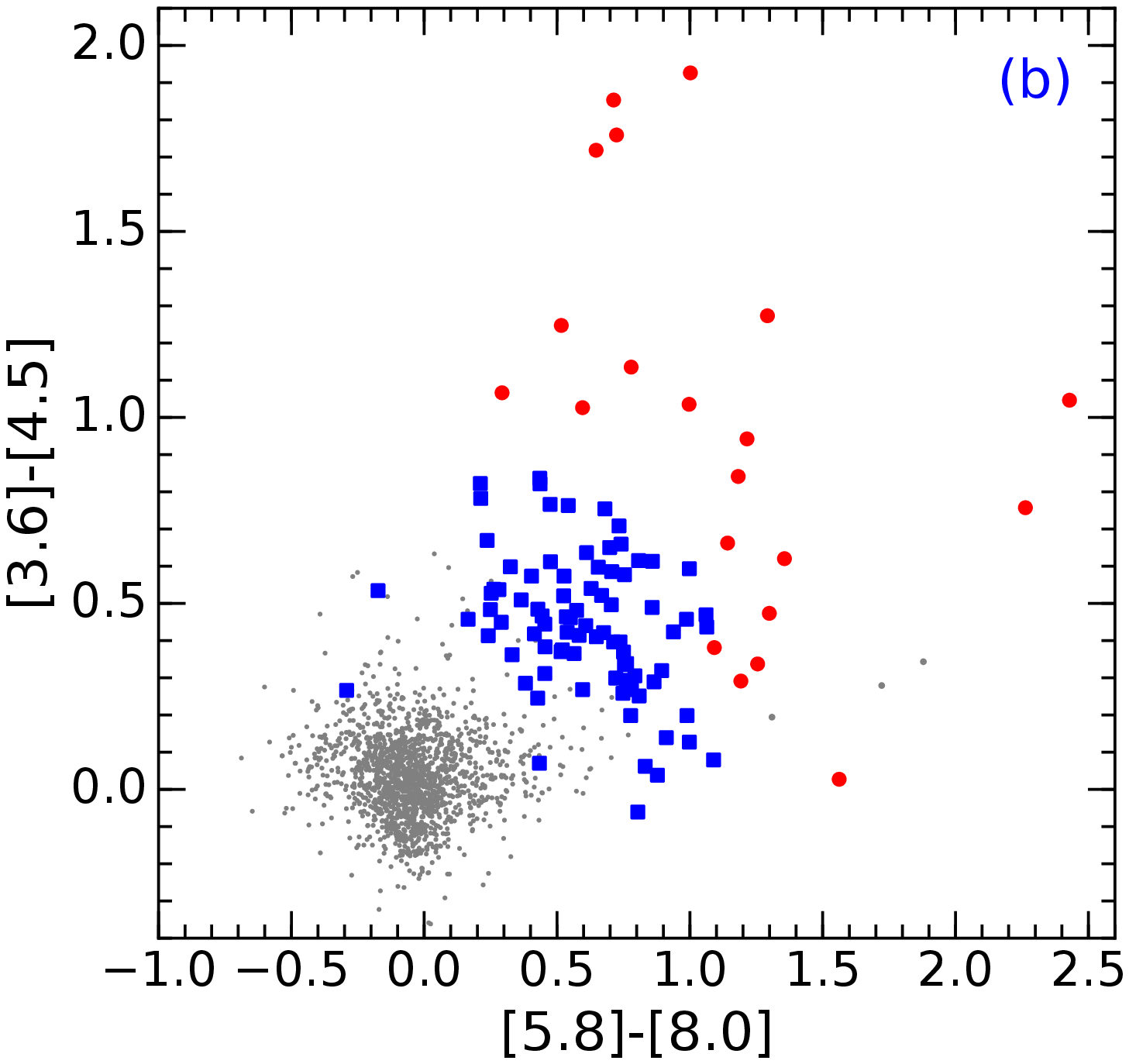} \\ \hfill
    \includegraphics[width=0.45\textwidth]{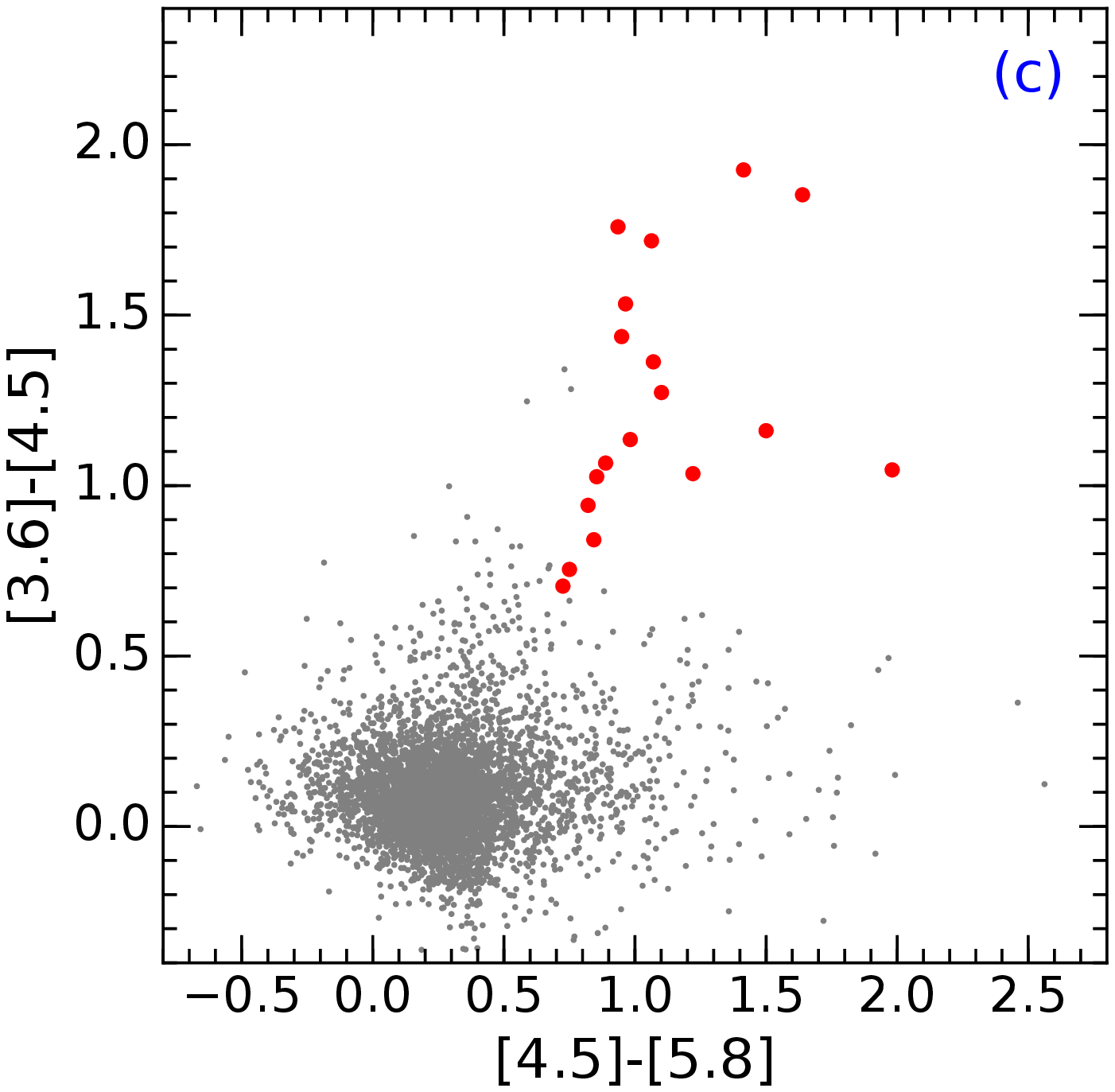} & \includegraphics[width=0.45\textwidth]{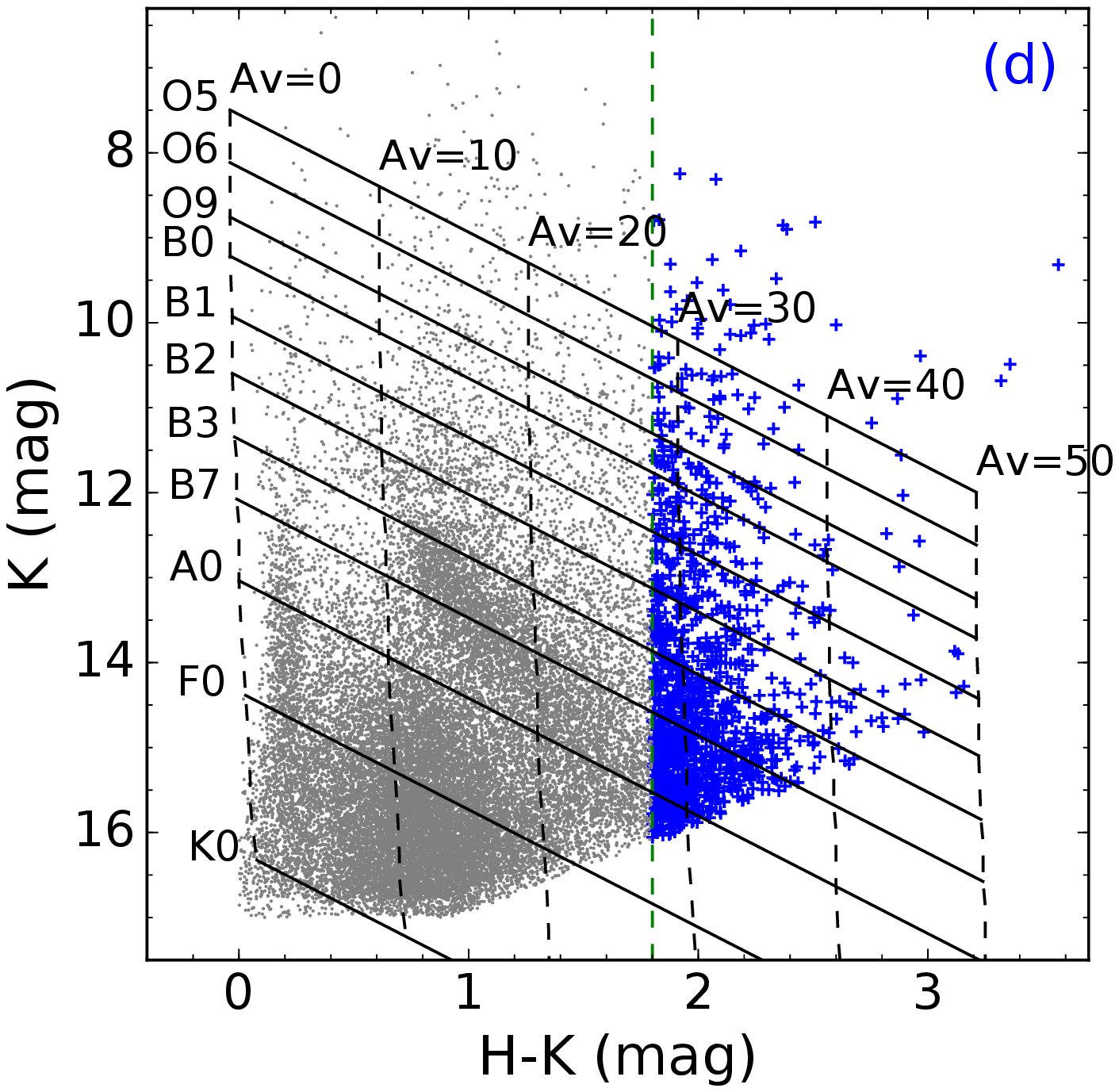} \\
%    \caption{}
\end{tabular}
%\epsscale{0.9}
%\plotone{fig7.eps}
\caption{\scriptsize YSO selection schemes using the MIR and NIR color-color and color-magnitude criteria. (a) Color-magnitude diagram ([3.6]$-$[24]/[3.6]) of the sources seen toward the N37 region. In the
 diagram, the selected sources are classified into different classes following the color criteria described in \citet{guieu10} and \citet{rebull11}. Sources at different classes are marked with different
 symbols: Class I (red circle), Flat spectrum (green pentagon), Class II (blue squares) and Class III (black triangles). An extinction vector for $A_K = $5 mag is also shown in the diagram. (b) Color-color
 diagram ([5.8]$-$[8.0]/[3.6]$-$[4.5]) of the sources that are detected in all the four {\it Spitzer}-IRAC bands. The sources are classified after removing possible contaminants based on the classification
 schemes of \citet{gutermuth09} (see text for more details). (c) Color-color diagram ([4.5]$-$[5.8]/[3.6]$-$[4.5]) of the sources detected in the first three {\it Spitzer}-IRAC bands (except 8 $\mu m$). 
 The color criteria for this classification were obtained from \citet{hartmann05} and \citet{getman07}. (d) Color-magnitude diagram (H$-$K/K) of all the sources detected in $H$ and $K$ bands. A cutoff H$-$K
 color value (i.e. 1.8 mag) was obtained from a nearby reference field ($15\arcmin \times 15\arcmin$ area centered at $l$ $=$ 25$^\circ$.372, $b$ $=$ 0$^\circ$.676).}
\label{fig7}
\end{figure*}
\subsubsection{Selection of YSOs}
Four different schemes are employed to identify and classify the YSOs for the selected 15$' \times$15$'$ area around the N37 region.

1. Young sources are known to be a strong emitter at MIR bands, while they are still embedded in their parent molecular clouds and cannot be seen in the optical/NIR bands. Hence, MIR
 photometric criteria allow us to identify sources at a very early phase. We cross-matched the sources that have detections in both the MIPSGAL 24 $\mu m$ and  {\it Spitzer}-IRAC/GLIMPSE
 3.6 $\mu m$ bands, and constructed a color-magnitude diagram ([3.6]$-$[24]/[3.6]) to identify the YSOs, following the color criteria given in \citet{guieu10} and \citet{rebull11}.
 A total of 100 sources are found that are common in the 3.6 and 24 $\mu$m bands. The color-magnitude diagram of these sources is shown in Figure~\ref{fig7}a. Different classes of YSOs are
 marked by distinct  symbols and are separated by black dashed-lines. The boundaries for other contaminants like disk-less stars and galaxies are also marked by a green dashed curve,
 following the criteria given in \citet{guieu10}. Using this scheme, a total of 19 Class I, 12 Flat-spectrum, 22 Class II, and 36 Class III sources were identified.

2. There are several sources that are not seen in the MIPSGAL 24 $\mu m$ image, but detected in the {\it Spitzer}-IRAC/GLIMPSE bands (3.6, 4.5, 5.8, and 8.0 $\mu m$). Hence, the color-color
 diagram ([5.8]$-$[8.0])/([3.6]$-$[4.5]) of the sources detected in all four IRAC bands was used to identify the additional YSOs (see Figure~\ref{fig7}b). Possible contaminants (such as
 broad-line active galactic nuclei, PAH-emitting galaxies and shock emission knots) were removed from the sample using the criteria given in \citet{gutermuth09}. The selected
 YSOs were classified into different evolutionary stages using the slopes of the {\it Spitzer}-IRAC/GLIMPSE spectral energy distribution (SED) (i.e. $\alpha_{IRAC}$) measured from 3.6 to
 8.0 $\mu$m \citep[see][for more details]{lada06}. Finally, using this scheme, we identified a total of 20 Class I and 99 Class II YSOs.

3. Due to prominent nebulosity seen in the IRAC 8.0 $\mu m$ band, there are several sources that are not detected in the 8.0 $\mu m$ image, but identified in other three GLIMPSE-IRAC
 bands (3.6, 4.5, and 5.8 $\mu m$). Hence, the color-color diagram ([3.6]$-$[4.5]/[4.5]$-$[5.8]) was constructed to identify the additional YSOs (see Figure~\ref{fig7}c). The sources that
 follow [4.5]$-$[5.8] $\geq$ 0.7 mag and [3.6]$-$[4.5] $\geq$ 0.7 mag are classified as protostars \citep{hartmann05,getman07}. Using this scheme, a total of 17 protostars were identified
 in the region around the bubble.

4. Presence of circumstellar material makes YSOs to appear much redder than the nearby field stars in the NIR color-magnitude diagram. Hence, we also used NIR color-magnitude diagram (H$-$K/K) to
 identify additional YSOs toward the N37 region. The $H-K$ color cut-off of 1.8 was estimated by constructing the color-magnitude diagram (H$-$K/K) of a nearby field region
 (size$\sim$15$' \times$15$'$ area centered at $l=$ 25$^\circ$.372; $b=$ 0$^\circ$.676). This color cut-off differentiate the field stars from the sources having large NIR excess. The
 NIR color-magnitude diagram of the sources is shown in Figure~\ref{fig7}d. Using the NIR scheme, we identified a total of 1203 red sources that could be presumed as YSOs.

There could be overlap of YSOs identified using these four different schemes. In order to have a complete catalog, YSOs identified using different schemes were cross-matched. Finally,
 a total of 29 Class I, 12 Flat spectrum, 99 Class II, 973 Class III YSOs, and 1066 red sources are identified toward the N37 region.
\subsubsection{Surface density analysis of YSOs}
\label{surface_density}
\begin{figure} 
\epsscale{1.2}
\plotone{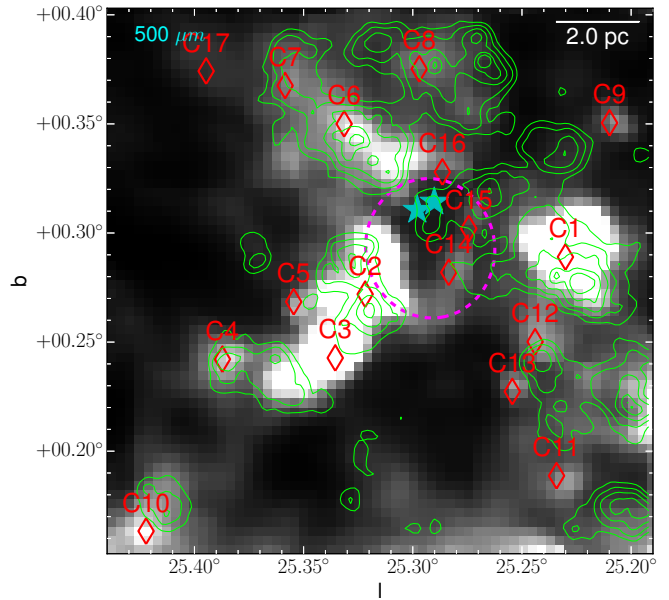}
\caption{\scriptsize Overlay of 20NN surface density contours (green) on the {\it Herschel} SPIRE 500 $\mu m$ image. The background 500 $\mu m$ image is similar to the one shown in Figure~\ref{fig2}. The surface density
 contours of YSOs are drawn at 15, 18, 22, 27, and 30 YSO pc$^{-2}$. The YSOs' clusters are seen toward the pillar, the IRDC, and the clump C1.}
 \label{fig8} 
\end{figure}

In order to examine how the YSOs are clustered in the region around the bubble, a nearest-neighbor (NN) surface density analysis of YSOs was performed following the method given in
 \citet{schmeja08} and \citet{schmeja11}. Using Monte Carlo simulations, \citet{schmeja08} showed that the 20NN surface density is capable to detect clusters with 10--1500 YSOs. Hence,
 the 20NN surface density analysis of YSOs was performed using a grid size of 6$\farcs$9 which corresponds to 0.1 pc at a distance of 3 kpc. Figure~\ref{fig8} shows the surface density
 contours of YSOs overlaid on the {\it Herschel} 500 $\mu m$ image. The contour levels are drawn at 15, 18, 22, 27, and 30 YSOs pc$^{-2}$. The positions of all the clumps
 identified in the {\it Herschel} column density map are also marked on the image. The YSO clusters are found toward the IRDC, the clump C1 and the pillar-like structure, and many of
 these YSO clusters are associated with peaks having more than 27 YSOs pc$^{-2}$ (see Figure~\ref{fig8}). It is already mentioned before that the clump C1 is part of the C25.29+0.31
 cloud and is not associated with the N37 molecular cloud. Therefore, the clusters of YSOs located towards the clump C1 might not have any physical association with the N37 bubble.
%
%Figure 8
%
\subsubsection{Spectral Energy Distribution of selected YSOs}
In order to infer the physical properties of YSOs (e.g., mass, age), the SED modeling of a few selected YSOs was performed using the SED fitter tool of \citet{robitaille06,robitaille07}.
 The grids of YSO models were computed using the radiation transfer code of \citet{whitney03a, whitney03b}, which assumes an accretion scenario for a pre-main sequence central star,
 surrounded by a flared accretion disk and a rotationally flattened envelope with cavities. The model grid has 20,000 SED models from \citet{robitaille06}, estimated using two-dimensional
 radiative transfer Monte Carlo simulations. Each YSO model gives the output SEDs for 10 inclination angles with masses ranging from 0.1--50 M$_{\odot}$. The fitter tool tries to find
 the best possible match of YSO models for the observed multi-wavelength fluxes followed by a chi-square minimization. The distance to the source and interstellar visual extinction
 (A$_{\rm V}$) are used as free parameters. We performed the SED modeling of those YSOs that have fluxes at least in five filter bands (among NIR JHK and {Spitzer}-IRAC bands), in order
 to constrain the diversity of the modeling parameters. Accordingly, a total of 81 YSOs were selected for the SED fitting. In the models, we used the A$_{\rm V}$ in the range from 0--50
 mag and the distance ranging from 2--4 kpc. For each YSO, only those models were selected which follow the criterion: $\chi^{2}$ -- $\chi^{2}_{best}$ $<$ 3, where $\chi^{2}$ is taken
 per data point. Note that the output parameters for each YSO are not unique because several models can satisfy the observed SED. Hence, the weighted mean values were computed for all
 the model fitted parameters for each YSO. In Table~\ref{table2}, we have listed, the right ascension (J2000), declination (J2000), the weighted mean values of the stellar age, stellar
 mass, total luminosity, extinction and evolutionary class for a sample of 6 YSOs. The complete table of 81 YSOs is available online in machine readable format.
%
%Table 2
\begin{deluxetable*}{llccccl}
\tablewidth{0pt}
\tablecaption{Physical parameters of selected YSOs derived from the SED modeling. \label{table2}}
\tablehead{
\colhead{RA (J2000)}&\colhead{Dec (J2000)}&\colhead{Log (Age)}&\colhead{Mass}       & \colhead{log (L$_{\rm tot}$)} & \colhead{A$_{\rm V}$} & \colhead{Class}\\    
\colhead{(hh:mm:ss)}&\colhead{(dd:mm:ss)} &\colhead{(yr)}     &\colhead{(M$_\odot$)}& \colhead{(L$_\odot$)}         & \colhead{(mag)}       & \colhead{ }}
\startdata
18:35:58.8     & -06:43:18 & 5.52$\pm$0.51 & 2.11$\pm$1.00 & 1.70$\pm$0.16    & 11.49$\pm$4.41    &   Class II \\  
18:36:17.7     & -06:45:41 & 5.55$\pm$0.72 & 1.54$\pm$1.32 & 1.62$\pm$0.24    &  3.74$\pm$2.18    &   Class II \\  
18:36:20.0$^C$ & -06:44:21 & 4.84$\pm$0.55 & 2.17$\pm$1.26 & 1.59$\pm$0.19    & 33.99$\pm$15.19   &   Class I  \\  
18:36:24.9$^B$ & -06:39:41 & 6.37$\pm$0.42 & 4.97$\pm$0.91 & 1.81$\pm$0.13    &  3.86$\pm$2.01    &   Class II \\  
18:36:25.5$^B$ & -06:39:46 & 5.48$\pm$0.60 & 3.94$\pm$0.91 & 1.53$\pm$0.28    &  0.73$\pm$1.12    &   Class I  \\  
18:36:36.1$^P$ & -06:37:51 & 5.64$\pm$0.43 & 2.22$\pm$0.98 & 1.66$\pm$0.20    &  2.63$\pm$1.50    &   Class II  
\enddata
\tablenotetext{B}{YSOs located toward the N37 bubble}
\tablenotetext{C}{YSOs located toward C1}
\tablenotetext{P}{YSOs located toward the pillar}
\end{deluxetable*}
\subsection{Near-infrared H-band polarization}
The polarization of background starlight is often used to study the  projected plane-of-the-sky magnetic field morphology. The polarization vectors of background stars allow to trace the
 field direction in the plane of the sky parallel to the direction of polarization \citep{davis51}. 

\begin{figure} 
\epsscale{1.1}
\plotone{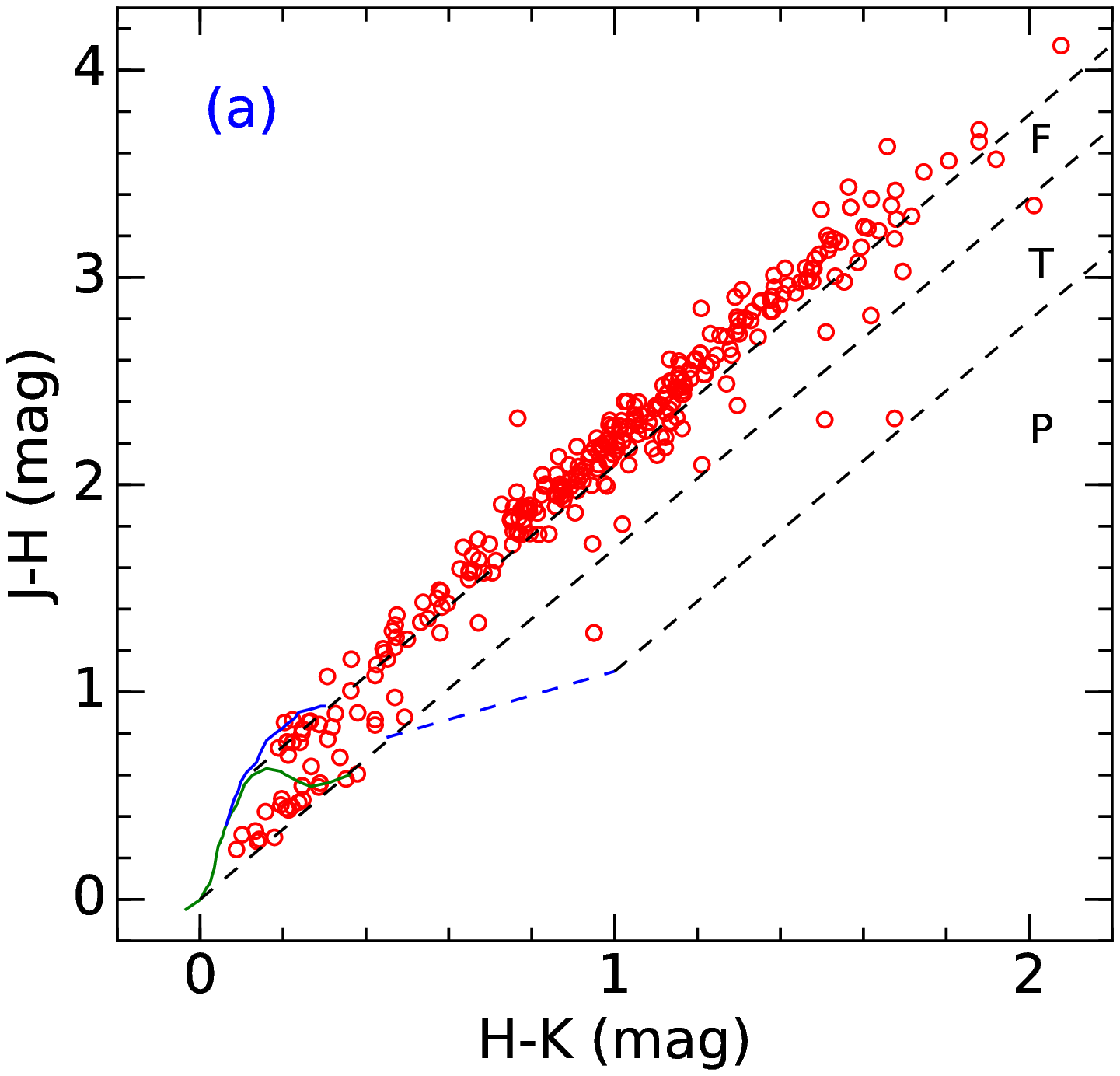}
\epsscale{1.1}
\plotone{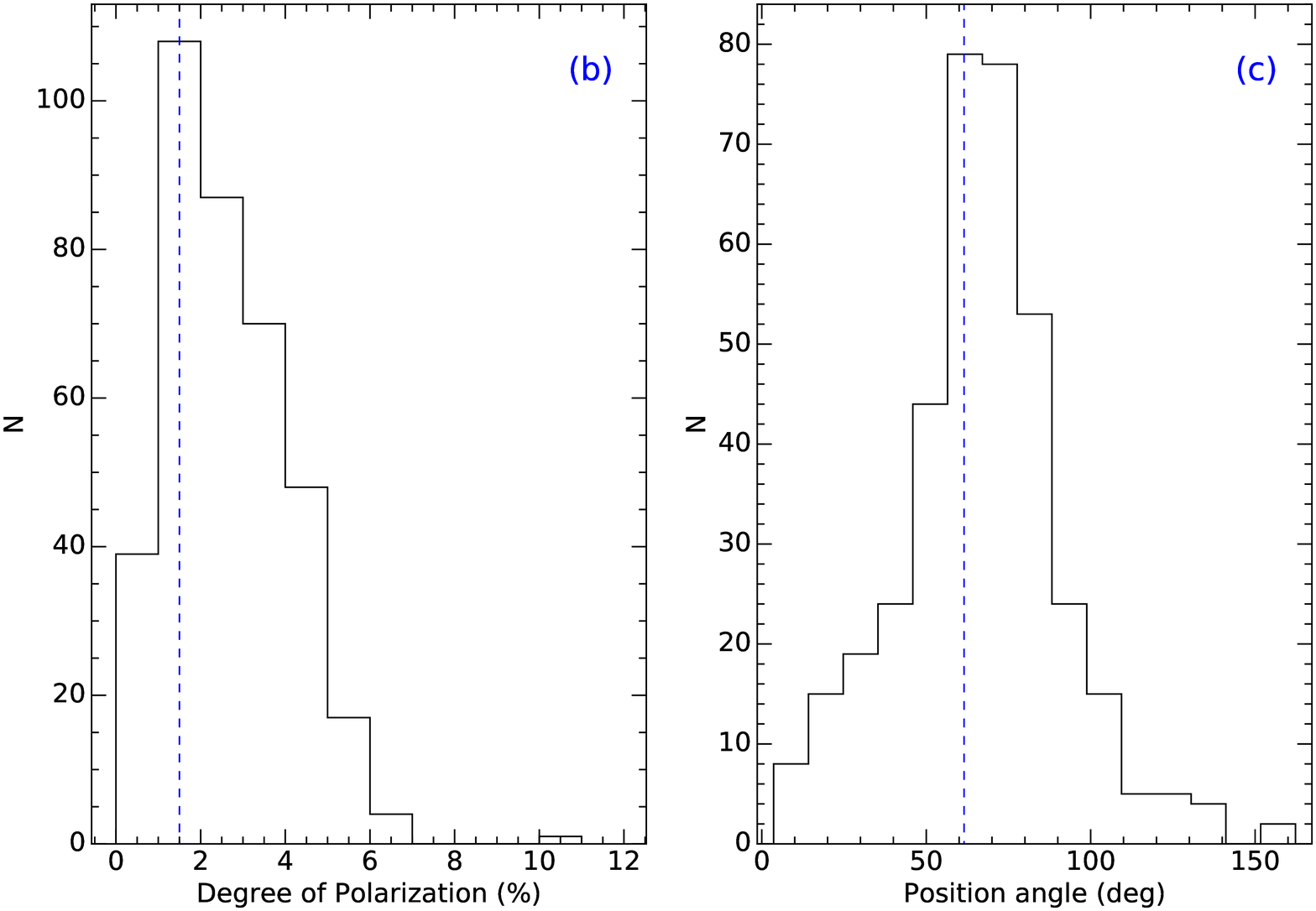}
\caption{\scriptsize (a) NIR color-color diagram (J-H/H-K) of the GPIPS sources with P/$\sigma_P \ge$ 2.5. The solid green and blue curves are the unreddened loci of main-sequence
 dwarf stars and giants \citep[from][]{bessell88} respectively. The blue dashed line shows the locus of Classical T Tauri (CTTS) obtained from \citet{meyer97}. Three
 parallel black dashed lines are the reddening vectors drawn from the base of the main-sequence locus, from the turning point of the main-sequence locus, and from the
 tip of the CTTS locus, using the reddening laws from \citet{cohen81}. The sources in the `F' region are generally evolved field stars, while the sources in the `T' region
 are mainly Class II YSOs, and the sources in the `P' region are considered as Class I YSOs \citep[see][for more details]{ojha04}. The color-color diagram
 implies that sources with $P/\sigma_P \ge$2.5 are mainly reddened background main-sequence and giant stars. (b) Histogram of the degree of polarizations and (c) histogram
 of the Galactic position angles of all the sources with $P/\sigma_P \ge$2.5 toward the N37 region.}
\label{fig9}
\end{figure}

NIR polarization data of point sources towards the N37 region were obtained from the GPIPS \citep[see][for more details]{clemens12} and were covered in several fields i.e., GP0608, GP0609,
 GP0610, GP0622, GP0623, GP0624, GP0635, GP0636, GP0637, GP0649, GP0650, GP0651. A total of 375 sources with reliable polarization measurements were identified in the 15$'\times$15$'$ area
 towards the N37 region using the criteria of P/$\sigma_P \ge$ 2.5 and UF of 1. In Figure~\ref{fig9}a, we show a color-color diagram ($H-K$ vs. $J-H$) of the selected sources
 and find that the majority of the sources are either reddened giants or main-sequence stars. Hence, it is evident from the NIR color-color diagram that the majority of stars are located
 behind the N37 molecular cloud. The histograms of the degree of polarization and the corresponding Galactic position angles are also shown in Figures~\ref{fig9}b and~\ref{fig9}c, respectively,
 which show that the majority of the sources have degree of polarization and position angle of about 1.5\% and $\sim$60$\degr$, respectively. If it is considered that the dust components
 responsible to polarize the background starlight are aligned along the magnetic field lines, then the corresponding plane-of-sky component of the magnetic field is oriented at a position
 angle of $\sim$60$\degr$.

%Figure 9
\begin{figure}
\epsscale{1.1}
\plotone{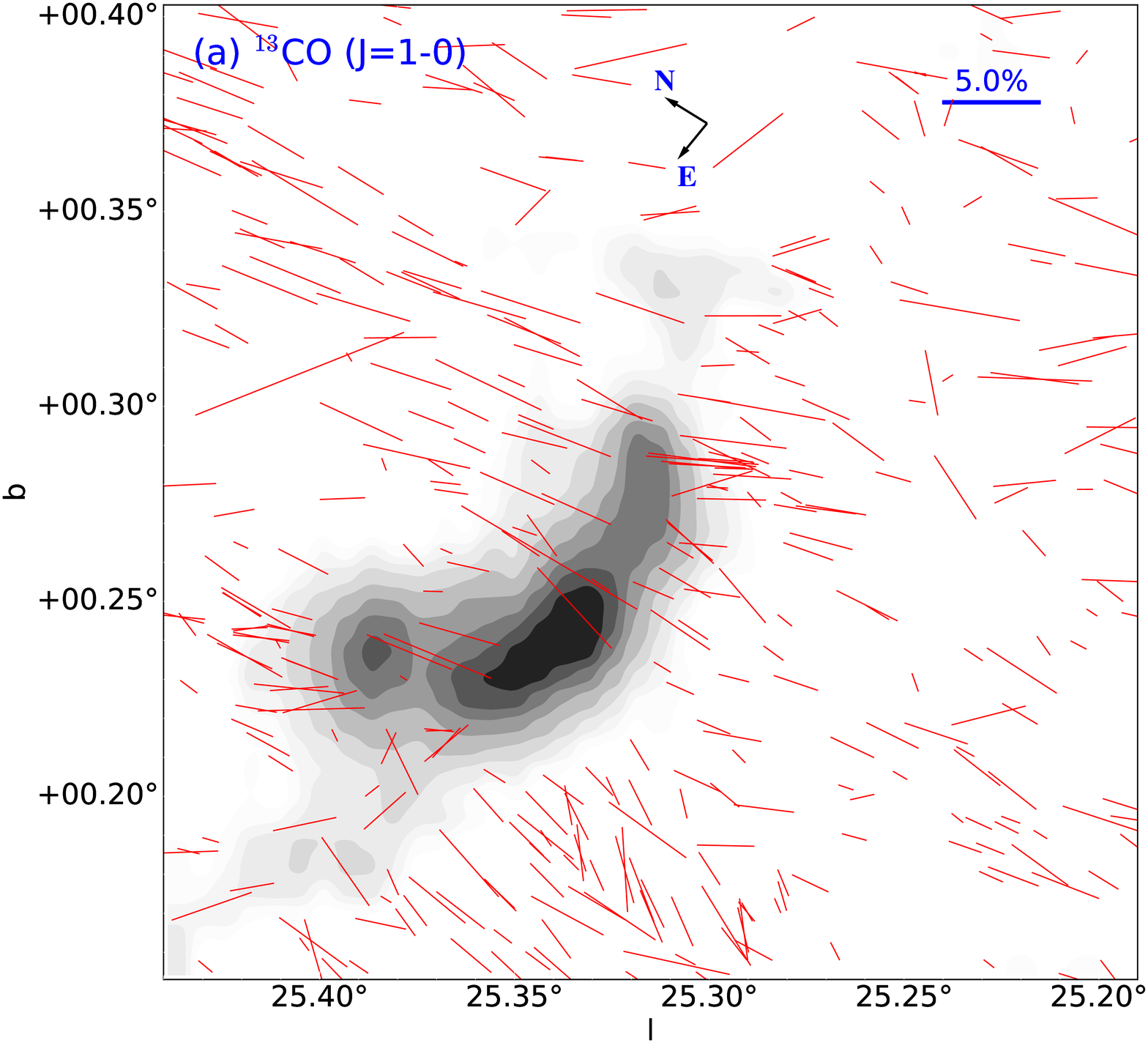}
\epsscale{1.1}
\plotone{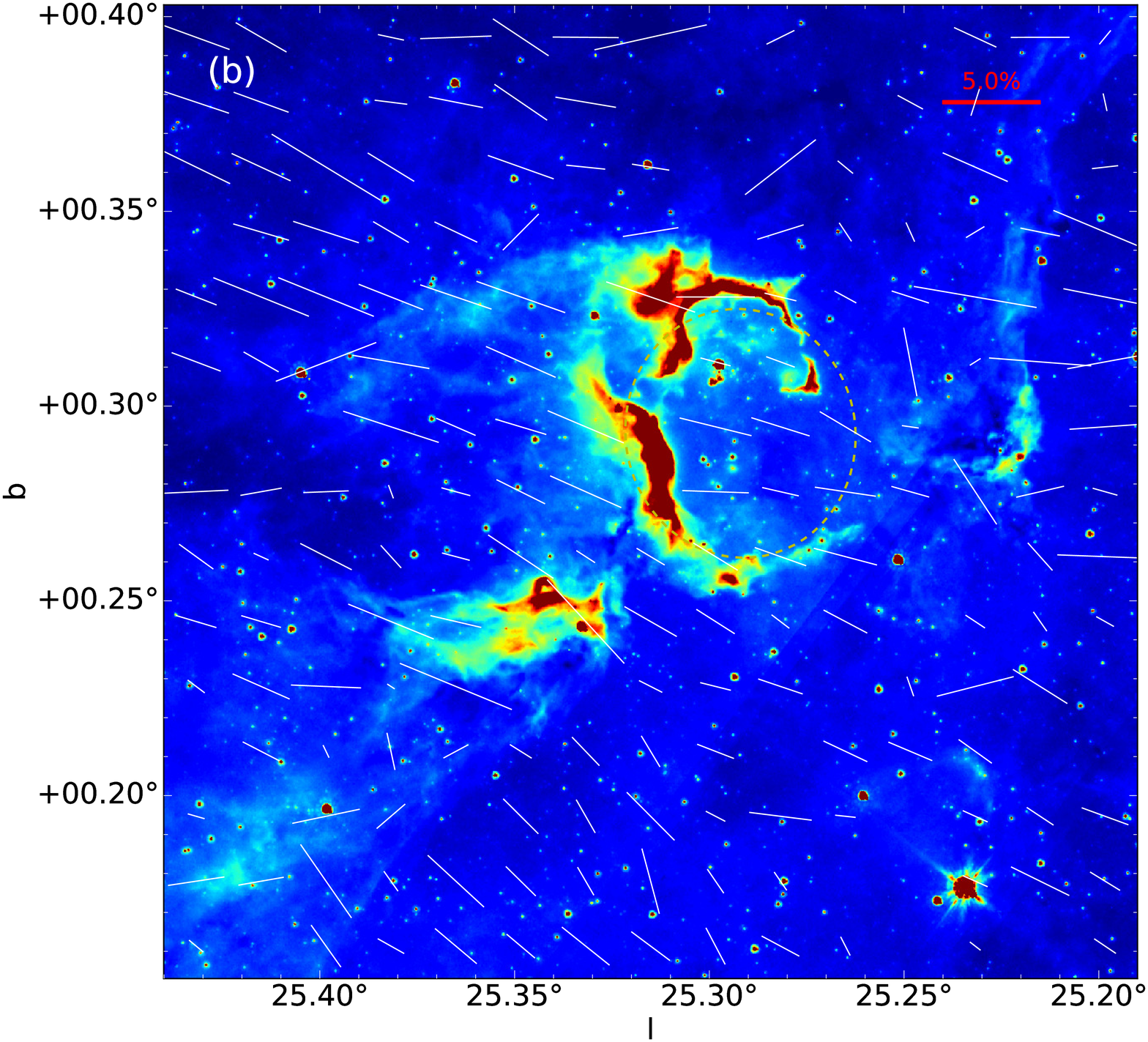}
\caption{\scriptsize (a) NIR $H$-band polarization vectors (red lines) of all the sources with P/$\sigma_P \ge$ 2.5, overplotted on the velocity integrated $^{13}$CO
 map. (b) Mean $H$-band polarization vectors (for a grid size of 1$' \times$1$'$) overplotted on the {\it Spitzer} 8 $\mu m$ image.}
\label{fig10}
\end{figure}

The polarization vectors overlaid on the velocity integrated $^{13}$CO map are shown in Figure~\ref{fig10}a. To examine the average distribution of NIR $H$-band polarization, the mean degree
 of polarization vectors superimposed on the {\it Spitzer} 8 $\mu m$ image are shown in Figure \ref{fig10}b. In order to study the mean polarization, our selected 15$'\times$15$'$ spatial
 area was divided into 225 grids having 1$'\times$1$'$ area for each grid and the mean polarization value for each grid is computed using the average Q and U Stokes parameters of all the
 $H$-band sources located inside that particular grid. Using the $H$-band polarization data, one cannot trace the morphology of the plane-of-the-sky projection of the magnetic field toward
 the dense clumps, where extinction is generally high enough for a background source to be detected in the NIR $H$-band.
%
%Figure 10
\subsection{Distribution and kinematics of molecular gas}
\label{sec:CO}
\begin{figure*}
\epsscale{0.9}
\plotone{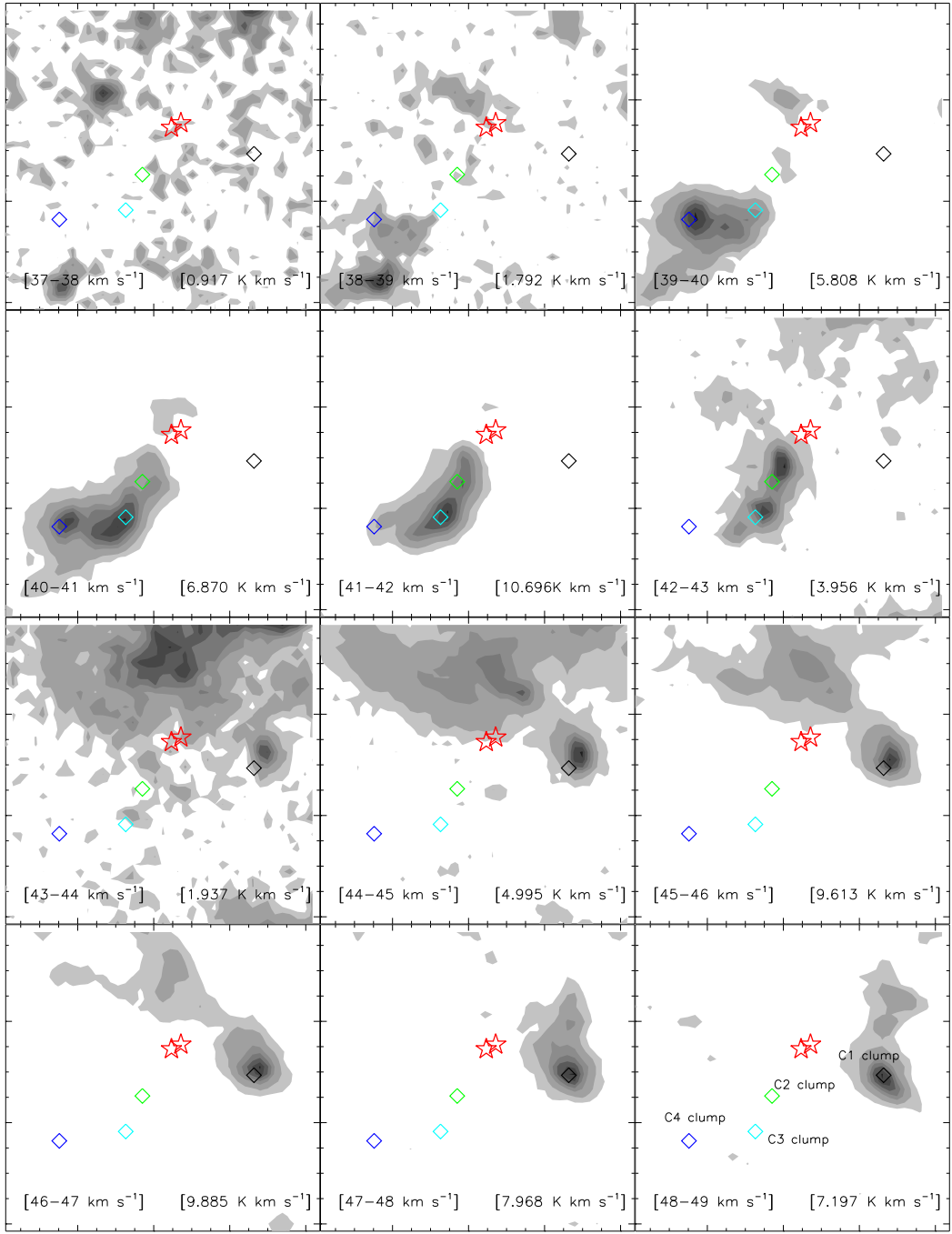}
\caption{\scriptsize The $^{13}$CO(J =1$-$0) velocity channel contour maps of the N37 region. The velocity range (in km s$^{-1}$) is indicated in the bottom left corner of
 each panel. The contours are drawn at 10, 20, 40, 55, 70, 85 and 98\% of the peak value, which is also written in the bottom right corner of each panel. Other marked symbols
 and labels are same as Figures~\ref{fig3} and \ref{fig6}.}
\label{fig11} 
\end{figure*}

The GRS $^{13}$CO (J=1--0) line data were utilized to examine the distribution and kinematics of the molecular gas towards the N37 bubble. Additionally, the velocity information of gas inferred
 from the $^{13}$CO data is used to know the physical association of different subregions seen in the selected region around the bubble. The integrated GRS $^{13}$CO (J=1$-$0) velocity channel maps
 (at intervals of 1 km s$^{-1}$) are shown in Figure~\ref{fig11}, tracing different subregions along the line of sight. As mentioned before from the $^{13}$CO profile that the molecular cloud
 associated with the bubble N37 (i.e., N37 molecular cloud) is depicted in the velocity range from 37--43 km s$^{-1}$. In this velocity range, three condensations (see last panel of
 Figure~\ref{fig11} for C2, C3, and C4) located towards the pillar-like structure are well detected in the channel maps. The integrated GRS $^{13}$CO intensity map for the
 C25.29+0.31\footnote[2]{http://www.bu.edu/iar/files/script-files/research/hii\_regions/region\_pages/C25.29+0.31.html} was previously reported by \citet{anderson09} having V$_{lsr}$ of
 $\sim$45.9 km s$^{-1}$ with a velocity range of $\sim$43--48 km s$^{-1}$. Based on the CO velocity profile and channel maps (Figure~\ref{fig11}), we infer that two nearby but distinct molecular
 clouds (i.e. N37 molecular cloud and C25.29+0.31) are present in our selected area of analysis. Hence, there could be a possibility for physical interaction between these two nearby molecular clouds.

%Figure 11
\begin{figure*}
\epsscale{0.9}
\plotone{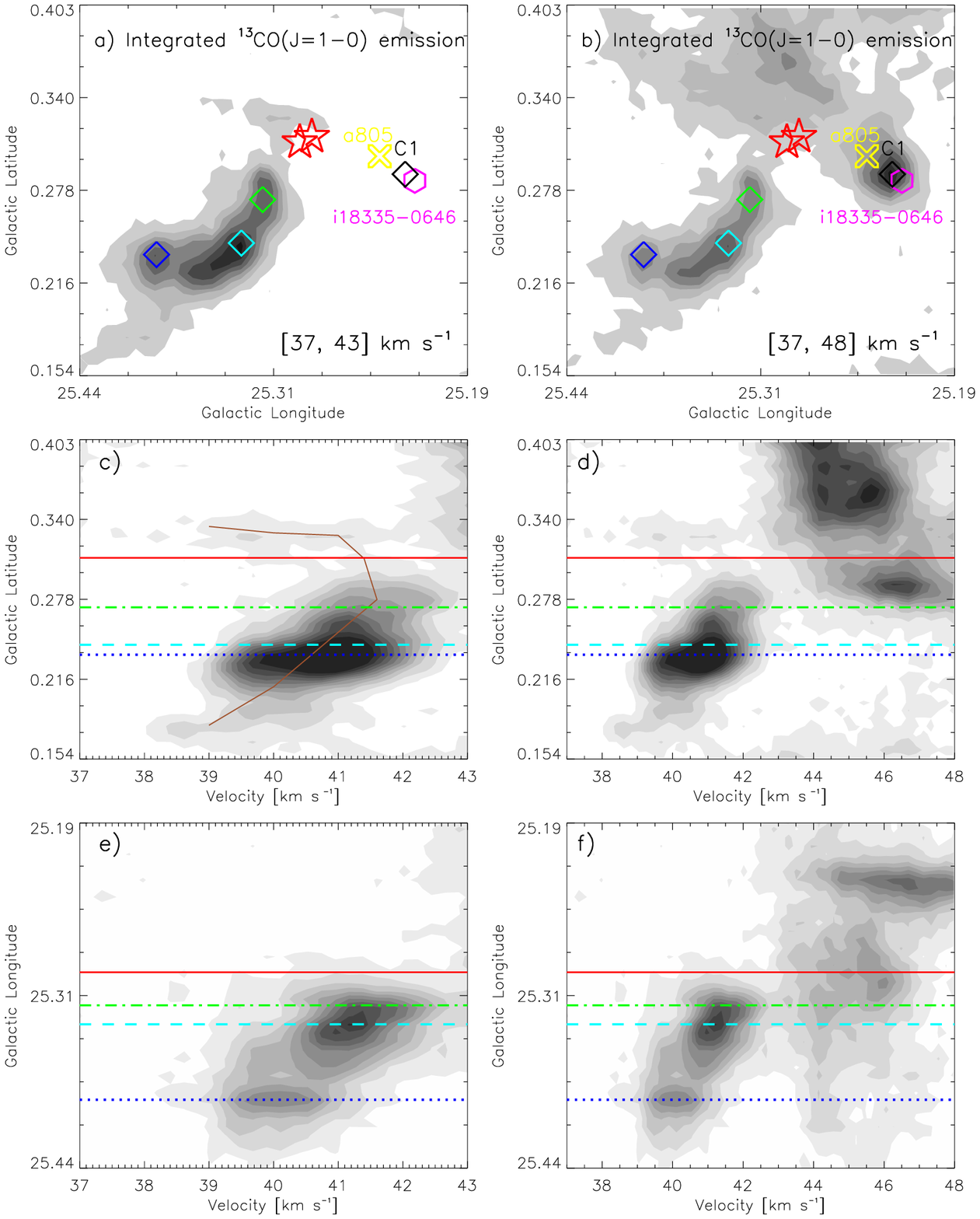}
\caption{\scriptsize The distribution of molecular gas toward the N37 molecular cloud and the cloud associated with IRAS 18335$-$0646/C1 clump. 
a) A contour map of integrated $^{13}$CO emission in the velocity range of 37 to 43 km s$^{-1}$. 
b) A contour map of integrated $^{13}$CO emission in the velocity range of 37 to 48 km s$^{-1}$. In the top two panels, four largest condensations seen in the {\it Herschel}
 column density map are also marked by diamond symbols. The positions of IRAS 18335$-$0646 (hexagon) and an O7II type source (a805; marked by an X symbol) are also highlighted
 in the top two panels. 
c) Latitude-velocity map of $^{13}$CO in the velocity range of 37 to 43 km s$^{-1}$. An inverted C-like structure is evident and is marked by a brown curve. 
d) Latitude-velocity map of $^{13}$CO in the velocity range of 37 to 48 km s$^{-1}$. 
e) Longitude-velocity map of $^{13}$CO in the velocity range of 37 to 43 km s$^{-1}$. 
f) Longitude-velocity map of $^{13}$CO in the velocity range of 37 to 48 km s$^{-1}$.
The molecular cloud associated with the bubble is traced in the velocity range of 37 to 43 km s$^{-1}$. The cloud associated with the IRAS 18335$-$0646/C1 clump is
 depicted in the velocity range of 43 to 48 km s$^{-1}$. The position-velocity diagram suggests the possibility of interaction between the N37 molecular cloud and
 the cloud associated with the C1 clump region (see text for more details). In all the position-velocity plots, the positions of ionizing star and condensations
 (associated with the N37 cloud) are shown by broken lines.} 
\label{fig12}
\end{figure*}

Note that the position-velocity analysis of these clouds is not yet explored. Figure \ref{fig12}a shows an integrated velocity map (37--43 km s$^{-1}$) of the region around the bubble N37,
 which reveals the physical association of molecular condensations with the N37 molecular cloud. In general, the position-velocity plots of the molecular gas are often used to search for any
 expansion of gas and/or outflow activity within a given cloud \citep[e.g.,][]{arce11,dewangan16}. The position-velocity diagrams of $^{13}$CO gas associated with the N37 cloud are shown in
 Figures~\ref{fig12}c and~\ref{fig12}e. The positions of the massive OB stars and the condensations are also marked in the position-velocity diagrams. The velocity gradients are evident toward
 the condensations C2, C3, and C4, which can be indicative of the outflow activities within each of them. Note that the angular resolution of the $^{13}$CO data (45$''$) is coarse therefore
 we cannot further explore the outflow activity within these condensations. Additionally, an inverted C-like structure appears in Figure~\ref{fig12}c (follow the marked curve) and the massive
 OB stars are located near the center of the structure. Such structure is indicative of an expanding shell associated with the H\,{\sc ii} region in the bubble N37 \citep[e.g.,][]{arce11,dewangan16}.
 The study of $^{13}$CO line data suggests the presence of molecular outflow(s) and the expanding H\,{\sc ii} region with an expansion velocity of $\sim$2.5 km s$^{-1}$. This expansion
 velocity corresponds to the half of the velocity range for the inverted C-like structure seen in the position-velocity diagram.

The integrated velocity map for a larger velocity range (37--48 km s$^{-1}$), which covers both the N37 and C25.29+0.31 molecular clouds, is presented in Figure~\ref{fig12}b. We have also
 constructed the position-velocity diagrams of $^{13}$CO gas in the corresponding velocity range (see Figures~\ref{fig12}d and~\ref{fig12}f). In Figures~\ref{fig12}d and~\ref{fig12}f, we find
 that the red-shifted component (43--48 km s$^{-1}$) and the blue-shifted component (37--43 km s$^{-1}$) are well separated by a lower intensity intermediated velocity emission, which is
 referred as a broad bridge feature. This feature in the position-velocity diagram is generally seen at the interface of the colliding molecular clouds \citep[see][for more detail]{haworth15a, haworth15b}.
 The implication of this feature is presented in the discussion section (Section~\ref{sec:discussion}).
%
%Figure 12
\section{Discussion}\label{sec:discussion}
As mentioned before, there are two molecular clouds (i.e. N37 molecular cloud and C25.29+0.31) present in the region around the bubble. The position-velocity analysis of the molecular gas
 towards these clouds reveals a broad bridge-like feature. This bridge-like feature is indicative of a cloud-cloud
 collision \citep{fukui14,haworth15a,haworth15b,torii15}. These authors also suggested that the collision between two molecular clouds can be a potential mechanism to trigger the formation of massive
 stars. Very recently, observational evidences of the cloud-cloud collision and the formation of massive stars through this process have been reported in the Galactic star-forming regions RCW120
 \citep{torii15} and RCW 38 \citep{fukui16}.

According to \citet{habe92} and \citet{torii15}, a collision between two non-identical clouds can produce a dense layer at the interface of these clouds and can create a cavity in the large cloud
 \citep[see Figure 12 of][]{torii15}. The compressed dense layer has the ability to develop the dense cores, which can subsequently form massive stars. After the formation of massive stars, their
 strong UV radiation can ionize the surrounding gas and develop an H {\sc ii} region. \citet{torii15} suggested that the cloud-cloud collision has formed an O-type star in the RCW 120 region, 
 and made it to appear as a broken bubble. It is mentioned before that the N37 molecular cloud hosts a pillar-like structure, IRDC, and the MIR bubble N37. The
 cloud also harbors YSOs clusters that are associated with the IRDC and pillar-like structure. It is possible that the cloud-cloud collision has influenced the star formation within the N37 molecular
 cloud (including the IRDC, OB stars, and the pillar-like structure). The massive OB stars might have formed due to a similar formation mechanism as it is reported for RCW 120. With time, the
 massive OB stars developed an H\,{\sc ii} region and the MIR bubble morphology appears to be originated due to the expansion of the photoionized gas (see Section~\ref{sec:ratt}). It is also possible
 that the collision between two molecular clouds might have developed the broken cavity and that is why the N37 bubble has appeared with a broken structure. However, we do not have enough observational
 evidences to firmly conclude about the origin of the broken feature of the N37 bubble.
 
\begin{figure*}
\epsscale{0.5}
\plotone{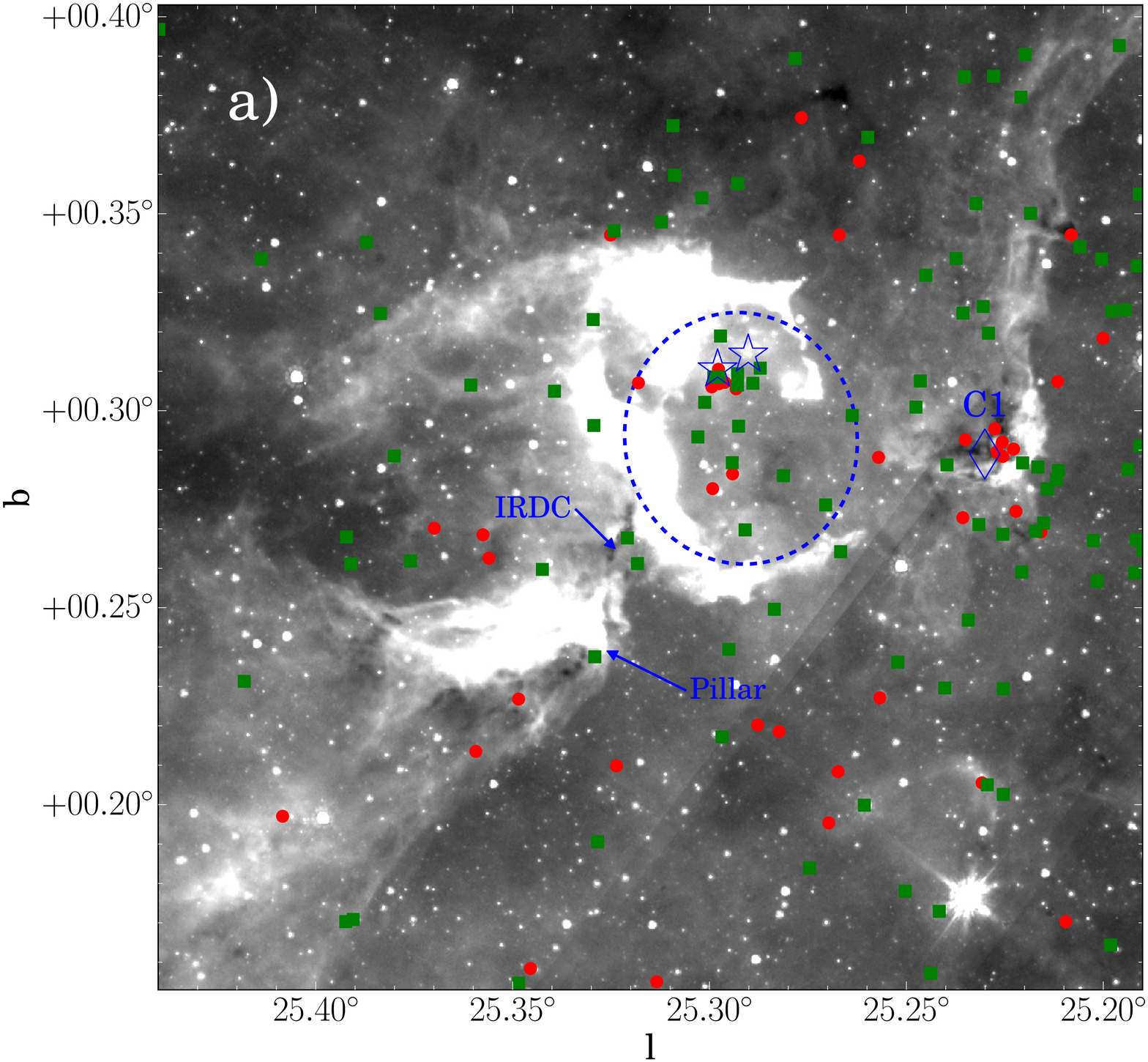}
\epsscale{0.5}
\plotone{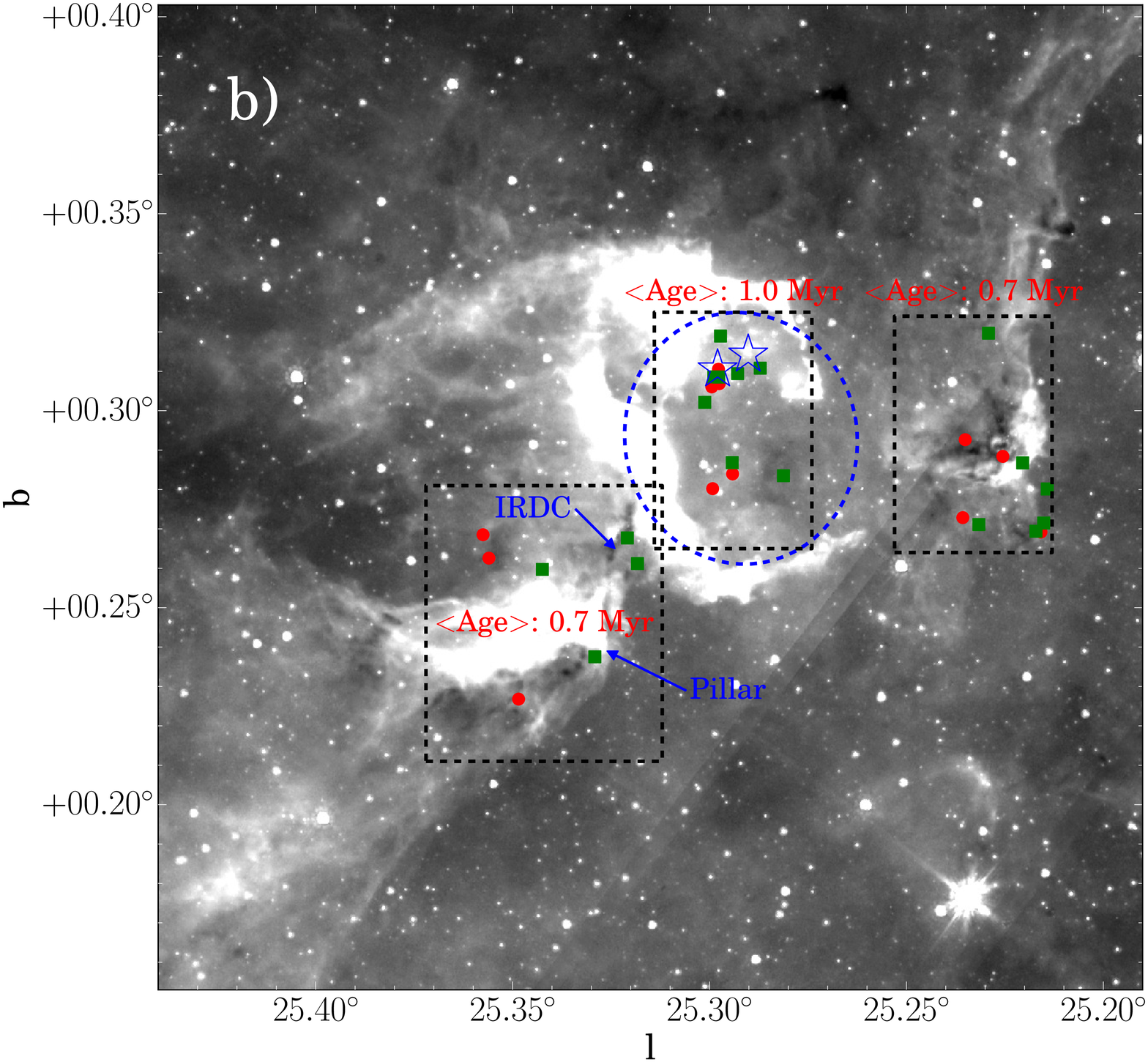}
\epsscale{0.6}
\plotone{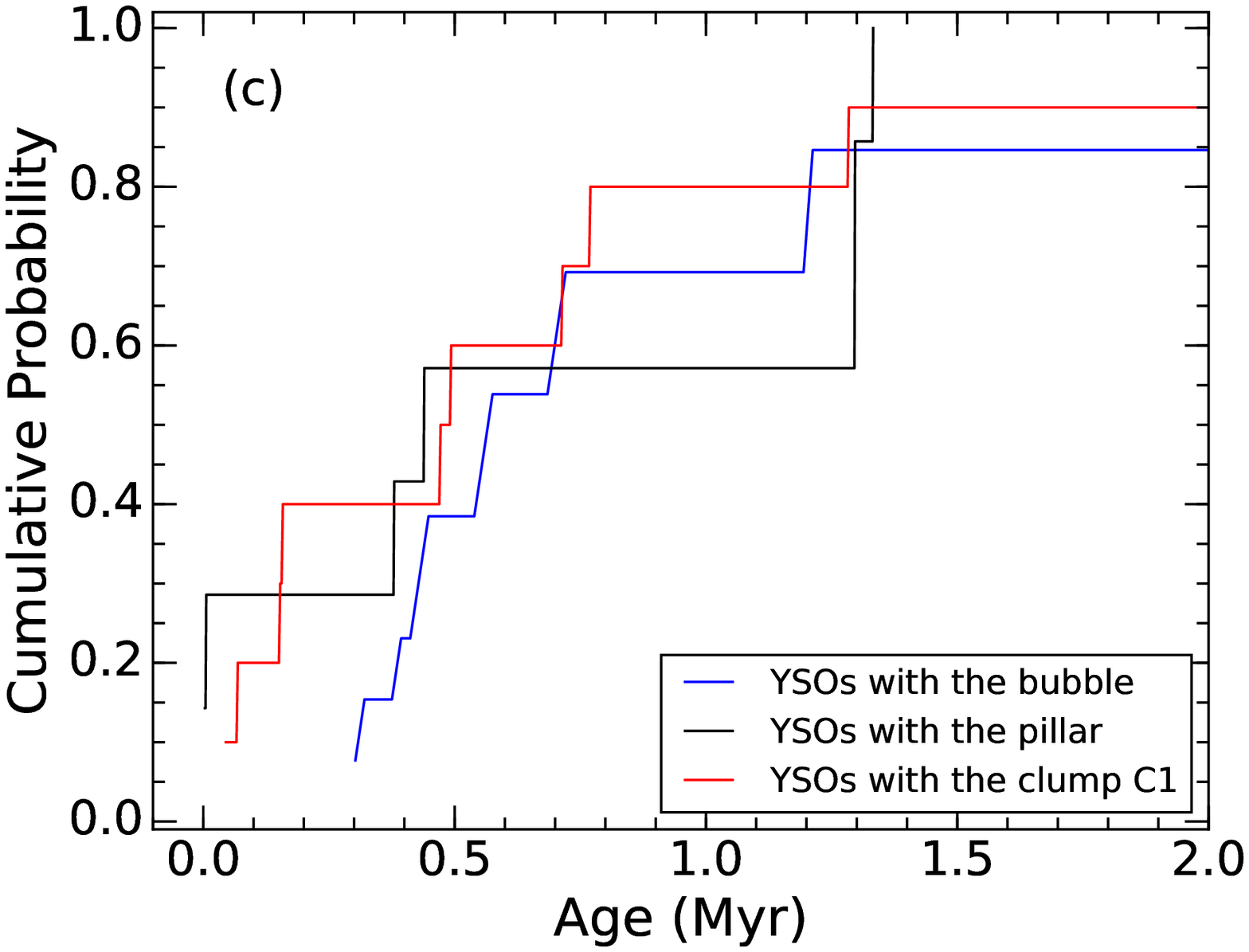}
\caption{\scriptsize (a) Class I (red circles) and Class II (green squares) YSOs are overplotted on the {\it Spitzer}-IRAC 8 $\mu m$ image of the N37 region. Other symbols are same
 as in Figure~\ref{fig1}. Some Class I YSOs are located inside the bubble and are also associated with C1 clump. A few Class I/II YSOs are also detected toward the
 pillar. (b) The positions of selected YSOs (in the N37 bubble, C1 clump, and pillars regions; see black dotted rectangles), for which the SED analysis was performed
 (see text for more details). In each selected region, the average age of YSOs is also shown. (c) Cumulative distributions of ages of YSOs toward the bubble, the pillar and
 the clump C1 are shown by blue, black and red lines, respectively.}
\label{fig13}
\end{figure*}

We calculated the dynamical age of the H\,{\sc ii} region (t$_{dyn}$) towards the N37 bubble to be $\approx$0.7 Myr for an ambient density of 10000 cm$^{-3}$ (see Section~\ref{sec:radio}).
 In general, the average ages of Class~I and Class~II YSOs are $\sim$0.44 Myr and $\sim$1--3 Myr \citep{evans09}, respectively. Considering these ages, it is unlikely that the star formation in the
 N37 cloud has been triggered by the expansion of the H\,{\sc ii} region. For further confirmation, we determined the average ages of YSOs toward the bubble, the pillar and the clump C1 which is,
 however, part of C25.29+0.31 molecular cloud. The distribution of Class I and Class II YSOs overplotted on the 8 $\mu m$ image are shown in Figure~\ref{fig13}a. A total of 13, 7 and 10 YSOs are
 found to be situated toward the N37 bubble, the pillar and the C1, respectively, for which the SED modeling was performed (see Figure~\ref{fig13}b). Corresponding mean ages of these YSOs are estimated
 to be 1.0, 0.7 and 0.7 Myr, respectively, which are comparable to the dynamical age of the H {\sc ii} region of 0.7 Myr. For a triggered star formation to occur the mean ages of YSOs should be less
 compared to the dynamical age of the H {\sc ii} region. However, the mean ages are associated with large standard deviations ($\sim$0.5 Myr) and hence, it is not possible to make a definite conclusion
 from this analysis.
 
We have also plotted the cumulative distribution of ages of YSOs toward the bubble, C1, and the pillar (see Figure~\ref{fig13}c). It can be seen in the cumulative distributions of the ages that
 the majority of the YSOs (at least $\sim$60\%) located towards the pillar are younger than the YSOs towards the bubble and C1. Hence, the formation of stars towards the pillar and C1 might have started later
 than the bubble. Though the YSOs associated with the pillar and the C1 are younger than the YSOs towards the bubble, the dynamical age of the H {\sc ii} region is not consistent enough to conclude whether
 they have formed due to the influence of bubble/massive OB stars. Note that in this paper we do not discuss results related to the C25.29+0.31 cloud, which hosts the clump C1, the IRAS 18335$-$0646,
 and the star a805 (an O7II spectral type) \citep[see][]{marco11}. It should be mentioned here that the pillars are generally assumed to be potential sights of triggered star formation
 \citep{klein80,elmegreen11}, and young stars are expected to appear at the tip of the pillars \citep{hester05}. A few Class I and Class II YSOs are found to be associated with the pillar (see
 Figure~\ref{fig13}a) and these YSOs might have formed by some other mechanisms than triggered by the OB stars or H {\sc ii} region.

It can be noticed in Figure~\ref{fig10}b, even though the mean polarization position angles are generally uniform at a Galactic position angle of $\sim$60$^o$ throughout the region, random changes
 in the polarization position angles are noticed near the interface of the N37 molecular cloud and the clump C1 which is part of the C25.29+0.31 molecular cloud. We suggest that this change in the
 polarization position angles can be explained by a distortion of gas due to the collision between these two molecular clouds.

%Figure 13
Overall, this region correlates well with the observational signatures proposed for the cloud-cloud collision process. In addition to the formation of OB stars, the cloud-cloud collision might have
 also triggered the formation of several other YSO clusters in the N37 molecular cloud.
\section{Conclusions}
\label{sec:conclusions}
We performed a multi-wavelength analysis of the Galactic MIR bubble N37 and its surrounding environment. The aim of this study is to investigate the physical environment and star formation mechanisms
 around the bubble. The main conclusions of this study are the following.

1. In the selected $15\arcmin  \times 15 \arcmin$ region around the MIR bubble N37, two molecular clouds (N37 molecular cloud and C25.29+0.31) are present along the line of sight. The molecular
 cloud associated with the bubble (i.e. N37 molecular cloud) is depicted in the velocity range from 37 to 43 km s$^{-1}$, while the C25.29+0.31 cloud is traced in the velocity range from 43 to 48 km
 s$^{-1}$. The N37 molecular cloud appears to be blue-shifted with respect to the C25.29+0.31 cloud.

2. Using photometric criteria, we find a total of seven OB stars within the N37 bubble, and spectroscopically confirmed two of these sources as O9V and B0V stars. The physical association of these
 sources with the N37 bubble is also confirmed by estimating their spectro-photometric distances. The O9V star is found as the primary ionizing source of the region. This result is in agreement
 with the Lyman continuum flux analysis using the 20 cm data.

3. Several molecular condensations surrounding the N37 bubble are identified in the {\it Herschel} column density map. The physical association of these condensations with the N37 bubble is inferred
 using the molecular gas distribution as traced in the integrated $^{13}$CO (J=1--0) map. Surface density analysis of the identified YSOs reveals that the YSOs are clustered toward these molecular condensations.

4. The mean ages of YSOs located in different parts of the region indicate that it is unlikely that these YSOs are triggered by energetics of the OB stars present within the bubble. This interpretation
 is supported with the knowledge of the dynamical age of the H {\sc ii} region.

5. The position-velocity analysis of $^{13}$CO data shows that two clouds (N37 molecular cloud and C25.29+0.31) are interconnected with a lower intensity emission known as broad bridge
 structure. The presence of such feature suggests the possibility of interaction between the N37 molecular cloud and the C25.29+0.31 cloud.

6. The position-velocity analysis of $^{13}$CO emission also reveals an inverted C-like structure, suggesting the signature of an expanding H\,{\sc ii} region. Based on the pressure calculations
(P$_{HII}$, $P_{rad}$, and P$_{wind}$), the photoionized gas associated with the bubble is found as the primary contributor for the feedback mechanism in the N37 cloud. Possibly
 the expanding H\,{\sc ii} region is responsible for the origin of the MIR bubble N37.

7. The collision between two clouds (i.e. N37 molecular cloud and C25.29+0.31) might have changed the uniformity of the molecular cloud which is depicted by a slight change in the polarization
 position angles of background starlight.

8. The collision between the N37 molecular cloud and the C25.29+0.31 cloud might have triggered the formation of massive OB stars. This process might also have triggered the formation of YSOs clusters in the
 N37 molecular cloud. 
\acknowledgments
We thank the anonymous referee for the constructive comments. We are also grateful to Dr. A. Luna for proving the IDL-based program for the analysis of CO line data. We thank the staff of IAO, Hanle and
 CREST, Hosakote, that made the observations using HCT possible. The facilities at IAO and CREST are operated by the Indian Institute of Astrophysics, Bangalore. We acknowledge the use of the AAVSO
 Photometric All-Sky Survey (APASS), funded by the Robert Martin Ayers Sciences Fund. This work is based on data obtained as part of the UKIRT Infrared Deep Sky
 Survey. This publication made use of data products from the Two Micron All Sky Survey (a joint project of the University of Massachusetts and the Infrared Processing and Analysis Center/ California
 Institute of Technology, funded by NASA and NSF), archival data obtained with the {\it Spitzer} Space Telescope (operated by the Jet Propulsion Laboratory, California Institute of Technology under
 a contract with NASA). This publication makes use of molecular line data from the Boston University-FCRAO Galactic Ring Survey (GRS). The GRS is a joint project of Boston University and Five College
 Radio Astronomy Observatory, funded by the National Science Foundation (NSF) under grants AST-9800334, AST-0098562, and AST-0100793. This publication makes use of the Galactic Plane Infrared
 Polarization Survey (GPIPS). The GPIPS was conducted using the {\it Mimir} instrument, jointly developed at Boston University and Lowell Observatory and supported by NASA, NSF, and the W.M. Keck
 Foundation. LKD is supported by the Department of Space, Government of India.
\end{document}